\documentclass[useAMS,usenatbib,usegraphicx]{mn2e}
\usepackage{aas_macros}
\usepackage[dvips]{graphicx}
\usepackage{longtable}
\usepackage{colortbl}   
\usepackage{lscape}
\usepackage{multirow}
\usepackage[figuresright]{rotating}
\usepackage{bigstrut}
\usepackage{amssymb}
\newcommand{\naco}{NaCo }

\title[Age and metallicity gradients in M87]{Age and metallicity gradients support hierarchical formation for M$87$\thanks{Based on European Southern Observatory (ESO) Very Large Telescope (VLT) programs 074.B-0404(A) and 076.B-0493(A) and observations made with the NASA/ESA \emph{Hubble Space Telescope}, obtained from the Data Archive at the Space Telescope Science Institute, which is operated by AURA, Inc., under NASA contract NAS 5-26555 }}
\author[M. Montes et. al.]{Mireia Montes$^{1,2}$\thanks{E-mail: mireia.montes.quiles@gmail.com (MM)}, Ignacio Trujillo$^{1,2}$, M. Almudena Prieto$^{1,2}$ and Jos\'{e} A. Acosta-Pulido$^{1,2}$ \\
$^{1}$Instituto de Astrof{\'i}sica de Canarias (IAC), V{\'i}a L{\'a}ctea s/n, La Laguna, E-38200, Spain\\ 
$^{2}$Departamento de Astrof{\'i}sica, Facultad de F{\'i}sica, Universidad de La Laguna, Astrof{\'i}sico Fco. S{\'a}nchez s/n, La Laguna, E-38207, Spain \\}

\begin{document}

\date{Accepted 2014 January 5. Received 2013 December 19; in original form 2013 August 7}
\pagerange{\pageref{firstpage}--\pageref{lastpage}} \pubyear{2013}
\maketitle

\label{firstpage}

\begin{abstract}

In order to probe the inside-out formation of the most massive galaxies in the Universe, we have explored the radial ($0.1\lesssim R \lesssim 8$ kpc) variation of the spectral energy distribution (SED) of M87 from UV to IR. For this purpose, we have combined high resolution data in $16$ different bands. Our analysis indicate that the age of the stellar population of M87 remains almost unchanged with radius. However, the metallicity ([Z/H]) profile presents three different zones: the innermost kpc shows a \textit{plateau} with supersolar metallicity, followed by a decline in metallicity down to $5$ kpc and another \textit{plateau} afterwards. The size of the inner \textit{plateau} is similar to the expected size ($R_e$) of an object with the predicted mass of M87 at $z=2$. The global [Z/H] gradient is $-0.26\pm0.10$, similar to those found in other nearby massive ellipticals. The observed change in the stellar population of M$87$ is consistent with a rapid formation of the central part (R$\lesssim5$kpc) of this galaxy followed by the accretion of the outer regions through the infall of more metal-poor material.
\end{abstract}


\section{Introduction}\label{intro}

Massive early-type galaxies are now widely believed to have been assembled hierarchically through repeated mergers which provide a relatively constant flow of accreted mass over a long time. The innermost regions of massive galaxies appear to have formed the majority of their stars at high redshift and on short timescales \citep[e.g.][]{Thomas2005} whereas their outer parts are likely assembled as a consequence of multiple major and minor merging \citep[e.g.][]{Trujillo2011}. This two-phase formation picture \citep[e.g.][]{Naab2009} agrees with the observed size evolution of the massive galaxies. At $z\sim 1$ ($2$) massive early-type galaxies ($M_*\sim10^{11}M_{\odot}$) were a factor of $2$ ($4$) smaller than present-day equal mass objects, having an average effective radii of only $\sim1$ kpc at $z\sim2$ \citep[e.g.][]{Daddi2005, Trujillo2006, Buitrago2008}. 

The merger, star formation and chemical enrichment history of massive galaxies are imprinted in their kinematics and chemical abundances. Consequently, measurements of these quantities on today massive galaxies should constrain their evolution and formation. Different formation scenarios predict different chemical enrichments for the massive galaxies. For instance, the classical model of monolithic collapse predicts that the inflowing gas is chemically enriched by evolving stars and contributes with metal-rich fuel for star formation. This inflowing gas gives rise to high metallicities in the galaxy centre and to significant logarithmic metallicity gradients that can be steeper than $-1.0$ and correlate strongly with galaxy mass \citep{Eggen1962,Larson1974,Carlberg1984}. However, revised monolithic models predict shallower metallicity gradients of $-0.3$ to $-0.5$ for a few massive galaxies \citep[e.g.][]{Chiosi2002, Pipino2008,Pipino2010, Merlin2012}. On the other hand, a hierarchical formation scenario, with major mergers, is expected to dilute the metallicity gradient \citep{Kobayashi2004}, reaching values of $-0.2$.

Many observational probes of stellar populations have been carried out for nearby early-type galaxies. Colours are one of the easiest observables for extragalactic systems. For instance, it has been known that the centres of low-resdhift ellipticals are redder that their outskirsts \citep[e.g.][]{Franx1990,Peletier1990}. These colour gradients are generally attributed to gradients in metallicity although a small contribution due to age gradients is expected \citep[e.g.][]{Spolaor2010}.
Works as \citet{Ogando2005}, \citet{Spolaor2010}, \citet{Kuntschner2010} and \citet{LaBarbera2010} show that logarithmic metallicity gradients in early-type galaxies are about $-0.2$ to $-0.3$ dex. In addition, \citet{Sanchez-Blazquez2006}, \citet{Kuntschner2010} and \citet{Loubser2012} found age gradients compatible with zero. However, evidences for a positive age gradient are also reported \citep[e.g.][]{Baes2007, LaBarbera2010}. There are a few but increasing number of studies exploring age and metallicity gradients in nearby massive galaxies up to several effective radii of radial distance \citep{Coccato2010, Tal2011, Roediger2011, Greene2012, LaBarbera2012}. These studies also agree on a metallicity decrease of the stellar populations of massive galaxies towards the outer regions as well as a non negligible contribution of younger stars \citep{LaBarbera2012}. As an example, \citet{Coccato2010} studied one of the brightest galaxies in the centre of the Coma Cluster, NGC~$4889$. They found evidences in 
its metallicity profile that points out to hierarchical formation. Two different gradients can be seen for R$\lesssim1.2R_e$ and R$\gtrsim1.2 R_e$. Furthermore, the abundance of [$\alpha$/Fe] remains flat down to $1.2R_e$ but decreases outwards. These evidences suggest that the outer parts of that galaxy were formed by accretion of low-mass systems, i.e. satellite galaxies.

In this paper, we concentrate our analysis on M$87$. M$87$ is an interesting target to explore the formation of massive galaxies. Located at the centre of the Virgo cluster, this object is probably one of the oldest massive galaxies in the Universe \citep{Kuntschner2010}. In this sense, its innermost region was most likely formed very early-on and consequently its stellar population should reflect the evolutionary path followed by this galaxy. \citet{Davies1993} provide absorption-line strengths for the centre of this galaxy reaching as far as $50\arcsec$ ($\sim4$ kpc). They found a negative gradient in $Mg_2$ and a positive gradient in $H_{\beta}$. However, the detected emission in $H_{\beta}$ could produce a dilution on this absorption line. As $Mg_2$ can be understood as a direct measurement of metallicity \citep[e.g.][]{Kuntschner2010}, their results point out to a negative gradient in metallicity. Furthermore, \citet{Kuntschner2010}, using SAURON data for the inner $100\arcsec$ ($\sim8$ kpc) 
of the galaxy, found no noticeable age gradient but a negative metallicity gradient. On the contrary, \citet{Liu2005} found a negative, but shallower gradient in metallicity and also a negative age gradient from $40\arcsec$ to $500\arcsec$. To sum up, all these evidences point out to a negative gradient in metallicity although the age gradient is much more difficult to determine. 

In this work, we take a step forward and explore with unprecedent resolution ($\sim0\farcs3$) the variation of the spectral energy distributions (SEDs) of M$87$ from $0.1$ kpc to $\sim 8$ kpc using $16$ different bands from the UV to the IR. The large wavelength range provided is crucial to obtain accurate estimates of age and metallicity \citep{Anders2004}. 

The vicinity of M$87$ and the high spatial resolution (HR) of the present work allow us to explore for the first time the substructure of the age and metallicity profiles. Well differentiated regions in the profiles could be an indication of different epochs of formation of this galaxy. Our paper will focus on this.

The structure of this paper is as follows. A description of the data and the photometry is presented in Section \ref{observations}. In Section \ref{secageandmet}, the colour (Section \ref{colours}), age and metallicity (Section \ref{SSPfitting}) gradients are derived and discussed. We discuss our results in Section \ref{discussion}. Finally, a summary is presented in Section \ref{summary}. The distance adopted for M$87$ is D=$16.1$ Mpc \citep[][i.e., at the distance of the galaxy, $1\arcsec=78 $ pc]{Blakeslee2001}. The photometry is given in the AB system unless indicated otherwise.

\section{Data}\label{observations}

\subsection{High resolution data}
The analysis conducted in this paper is based on multiwavelength coverage of the galaxy, the same as used in Montes et al. (2013, submitted) for the analysis of the globular cluster population of the central region of M87. A brief description of the data is presented here. Table \ref{datadetails} provides a description of the main characteristics of the dataset. The NIR data consist on HR images in the $J$ and $K_s$ bands acquired with the Very Large Telescope (VLT) using the Nasmyth Adaptive Optics System plus the Near-Infrared Imager and Spectrograh (NAOS + CONICA, NaCo, $0\farcs0271/\rm{pixel}$). The inner half square arcminute of M$87$ was observed in J and in $K_s$. The bright nucleus was used as reference for the adaptive optics. The achieved resolution was $0\farcs27$ ($J$) and $0\farcs19$ ($K_s$) measured as the full width at half maximum (\textsc{FWHM}) of the most compact source in each image. The \naco data reduction was performed using the \textsc{ECLIPSE} package provided by ESO. Photometric 
calibration of the images make use of standard stars taken along with the science frames. 
An image taken with NICMOS camera 2 using the filter F160W was also added from the \emph{HST} archive. Its field of view (FOV) is smaller ($19\farcs2\;\times\;19\farcs2$), so the photometry provided by this filter is limited to R$<10\arcsec$. This NIR dataset was complemented with \emph{HST} archival data covering the UV-optical range.  The spatial resolution achieved is $\sim0\farcs15$ for the ACS and $\sim0\farcs22$ for the WFPC2.

STIS far UV MAMA (F25SFR2) and near UV MAMA (F25QTZ) images of the central region of M$87$ were also extracted from the archive although the images only overlap with half of the \naco $J$-band field of view. The spatial resolution is $~0\farcs13$. The HST images were combined using \textsc{Multidrizzle} \citep{Fruchter2002} to achieve the nominal resolution of each camera. \textsc{Multidrizzle} combines the aligned individual exposures, rejects outliers and removes the geometric distortions. 

 \begin{table*}
  \centering
   \begin{tabular}{@{}lllcrccc@{}}\hline
Filter & $\lambda$ & $\Delta \lambda\,(\AA)$ & Instrument & Pixel scale ($\arcsec/pixel$) & Exp. Time ($s$) & Date & Program      \\ 
\hline
  F25SRF2    &            &              &                     &                 &           &               & \\
   FUV-MAMA  &  $1456.6$ & $120.8$       & STIS@\emph{HST}     &  $0.025$       &  $3526$   & $17/05/1999$  & $8140$ \\
  F25SRF2    &            &              &                     &                 &           &               & \\
   FUV-MAMA  &  $1456.6$ & $120.8$       & STIS@\emph{HST}     &  $0.025$       &  $2680$   & $14/02/2002$  & $8643$ \\ 
  F25QTZ     &            &              &                     &                 &           &               & \\
   NUV-MAMA  &  $2355.3$ & $419.8$       & STIS@\emph{HST}     &  $0.025$       &  $2372$   & $17/05/1999$  & $8140$ \\ 
  F336W      &  $3359.5$ & $204.5$       & WFPC2@\emph{HST}    &  $0.046$/$0.1$ &  $28800$  & $25/12/2000$  & $8587$  \\ 
  F410M      &  $4092.7$ & $93.8$        & WFPC2@\emph{HST}    &  $0.046$/$0.1$ &  $19200$  & $07/02/2001$  & $8587$  \\ 
  F467M      &  $4670.2$ & $75.3$        & WFPC2@\emph{HST}    &  $0.046$/$0.1$ &  $7200$   & $28/12/2000$  & $8587$ \\ 
  F475W      &  $4745.3$ & $420.1$       & ACS/WFC@\emph{HST}  &  $0.05$        &  $750$    & $19/01/2003$  & $9401$ \\ 
  F547M      &  $5483.9$ & $205.5$       & WFPC2@\emph{HST}    &  $0.046$/$0.1$ &  $7200$   & $10/02/2001$  & $8587$ \\
  F555W      &  $5442.9$ & $522.2$       & WFPC2@\emph{HST}    &  $0.046$/$0.1$ &  $2430$   & $03/02/1995$  & $5477$ \\
  F606W      &  $5919.4$ & $672.3$       & ACS/WFC@\emph{HST}  &  $0.05$        &  $28500$  & $24/12/2005$  & $10543$   \\ 
  F658N      &  $6590.8$ & $29.4$        & WFPC2@\emph{HST}    &  $0.046$/$0.1$ &  $13900$  & $09/05/1996$  & $6296$ \\ 
  F702W      &  $6917.1$ & $586.7$       & WFPC2@\emph{HST}    &  $0.046$/$0.1$ &  $280$    & $23/01/1995$  & $5476$ \\
  F814W      &  $8059.9$ & $653.0$       & ACS/WFC@\emph{HST}  &  $0.05$        &  $72000$  & $24/15/2005$  & $10543$  \\ 
  F850LP     &  $9036.4$ & $527.2$       & ACS/WFC@\emph{HST}  &  $0.05$        &  $1120$   & $19/01/2003$  & $9401$ \\ 
  $J$        &  $12650.0$ & $2500.0$     & NACO@VLT            &  $0.027$       &  $784$ & $23/01/2006$  & $076.B-0493(A)$ \\
  F160W      &  $16506.0$ & $1177.3$     & NICMOS2@\emph{HST}  &  $0.075$       &  $32$ & $20/11/1997$  & $7171$ \\
  $K_s$      &  $21800.0$ & $3500.0$     & NACO@VLT            &  $0.027$       &  $300$ & $23/01/2006$  & $076.B-0493(A)$ \\ \hline
  \end{tabular} 
   \caption{Set of filters used in this study of M87 with HR.}\label{datadetails}
 \end{table*}

\subsection{Photometric profiles}

\subsubsection{High resolution profiles of the centre of M$87$} \label{HRphot}

The HR surface brightness profiles of the centre of M87 in all filters were obtained using a custom-made task developed in IDL. The active nucleus and jet of M$87$ were identified and masked by eye while the detection of the globular clusters was made using \textsc{SExtractor} \citep{Bertin1996} and they were also masked. The central region of M$87$ is circular symmetric \citep{Cohen1986}, so the photometry was extracted using circular annuli at different radial distances. To unify the effects of the resolution in the different images, we take $3$ times the \textsc{FWHM} of the worst spatially resolved image ($J$) as the width of the rings for the photometry, i.e. $\sim1 \arcsec$. To avoid contamination due to the Point Spread Function (PSF) wings of the nuclear point source, the inner $1\arcsec$ was masked in all images. For each annulus, the surface brightness was obtained averaging the pixel values.

The prominent jet of M$87$ may have effects in the profiles of the galaxy. To test this possible source of contamination, we derived the surface brightness profiles at both sides around the nucleus: the right side, from the nucleus to the west, and the left side, from the nucleus to the east. The profiles of the right and left sides present the same properties within errors and the symmetry of the integrated properties of the galaxy are not affected by the presence of the extended component associated to the jet. 

Our HR photometry is given in Table \ref{tablarads} and the profiles presented in Figure \ref{figprof}.

\subsubsection{Wide-field profiles}

To explore the stellar populations down to M$87$'s effective radii \citep[$R_e=106\farcs2$, ][]{Falcon-Barroso2011}, the HR data described above was complemented with 2MASS\footnote{http://irsa.ipac.caltech.edu/cgi-bin/2MASS/LGA/nph-lga?objstr=m87} \citep{Jarrett2003}, SDSS $griz$ \citep{Chen2010} and GALEX \citep{GildePaz2007} surface brightness profiles collected from the literature. Typical spatial resolutions are: $\sim3\arcsec$ for 2MASS \citep{Skrutskie2006}, $\sim1\farcs5$ for SDSS \citep{York2000}, $\sim4\farcs5\textendash5\arcsec$ for GALEX \citep{GildePaz2007}. The exposure times of the images are: $7.8$s for 2MASS, $55$s for SDSS and $1585$s for GALEX. In addition, we have downloaded the $u$ band image from the SDSS archive to complement the profiles provided by \citet{Chen2010}.

We also present a brief description of the derivation of the surface brightness profiles of this wide-field dataset. This is important to ensure that these profiles are consistent with our HR profiles. The 2MASS profiles \citep{Jarrett2003} were derived using an almost circular annulus of axis ratio of $0.99$. Likewise, the GALEX mean surface brightness were derived within elliptical ($\epsilon=0.2$) annuli, as described in \citet{GildePaz2007}. \citet{Chen2010} used the IRAF task ELLIPSE to derive the SDSS $griz$ profiles, with variable ellipticity reaching a maximun of $0.13$ for R$\sim R_e$, while for $u$ we used circular annuli. The ellipticity values used to derive these profiles are very close to the circular apertures we used for the inner regions except in the case of the GALEX photometry. In any case, GALEX data will no be used to derive the main conclusions of this work as we will explain later on. The global extension of the radial profiles from the literature are: $240\arcsec$, $403\arcsec$ and $331\arcsec$, respectively. These profiles were used only up to $\sim100\arcsec$ for reasons explained in the following sections. SDSS and GALEX photometry were kindly provided by the authors.

\subsubsection{HR vs. wide-field photometry}

\begin{figure}
 \begin{center}
  \includegraphics[height=0.35\textheight]{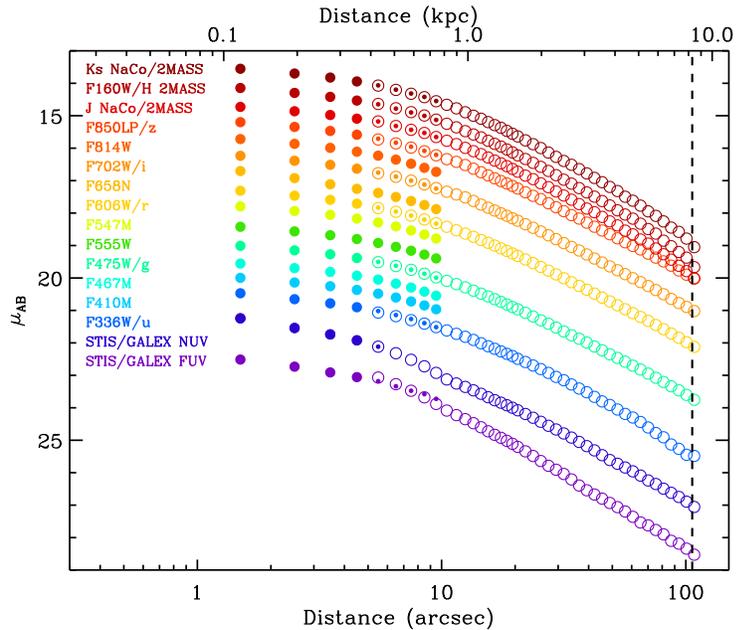} 
   \caption{Surface brightness profiles of the multiwavelength dataset for M$87$ up to $R_e$. The vertical dashed line marks the position of the $R_e$. Different vertical shifts to the profiles were applied for visualization. Filled circles correspond to HR photometry while open circles depict wide-field profiles from the literature.}\label{figprof}
 \end{center}
\end{figure}

The HR photometry, in Section \ref{HRphot}, describe the very centre of the galaxy (R$<1$ kpc). The field of view of the cameras used to explore the inner region of M$87$ is entirely dominated by the flux of the galaxy as it is observing its inner regions. This results in an impossibility to obtain a proper sky background for these images. Therefore, the outer shape of the surface brightness profiles of the HR is compromised. To surpass this problem, we made use of wide field photometry from the literature to determine down to which distance the radial profiles of the HR images are reliable. This is illustrated in Figure \ref{figprof}, where the surface brightness profiles are shown. The HR profiles are plotted down to the radial distances where they are reliable. Filled circles correspond to HR photometry while open circles depict wide-field profiles from the literature. These profiles indicate which filter was used at a certain radius. To show the agreement of the HR and the outer region photometry, the profiles shown in Figure \ref{figprof} are shifted vertically for visualization purposes. As seen in Figure \ref{figprof} the HR profiles in the optical bands reach down to $10\arcsec$. For STIS F25QTZ, the distance reduces to $5\farcs5$. Outwards, GALEX, SSDS and 2MASS photometry were used down to M$87$'s $R_e$. 

To avoid zeropoint biases among the different data, we decided to rely our analysis on the ground-based photometry and shift the HST profiles. In order to estimate the offsets between HR and wide-field profiles, we have compared the profiles at $\sim10\arcsec$, to minimize the contribution of the PSF of the ground-based telescopes and the sky background oversubstraction in the HST bands. The offsets were calculated as the difference between the HST filters and the corresponding (or the nearest) ground-based filter. We provided the offsets and the root mean square (r.m.s) between HR and wide-field profiles in a common overlapping region (from $5\arcsec$ to $10\arcsec$) in Table \ref{app:tablashift} in Appendix \ref{appendixa}. In this overlapping region, the HR and wide-field data were used cojointly to build and analyse the SEDs, i.e. when estimating the $\tilde{\chi}^2$. This common HR and wide-field region is depicted by the superposition of open circles to the (smaller) filled, see Figure \ref{figprof}.

The photometry was corrected for Galactic reddening using the maps of \citet{Schlegel1998} ($E(B-V)=0.022$) and the \citet{Cardelli1989} extinction law. The presence of almost no internal dust in this galaxy, as shown in \citet{Baes2010}, indicates that internal dust extinction should not affect the results presented here.

\subsubsection{Photometric uncertainty estimation}\label{errors}

Obtaining a realistic error estimate of our photometry is crucial in order to constrain properly the age and metallicity of the stellar populations. As mentioned before, in the innermost region of M$87$ we have derived the photometric profiles using circular annuli at different radial distances and averaging the fluxes within the annuli. The nominal photometric uncertainty obtained in such way is extremely low ($<0.0003$ mag) and does not reflect the real precision that can be expected. The two main sources of error are: the uncertainty in the photometric zeropoint and the photometrical uncertainties. Consequently, to get a much better estimation of the photometric errors, for each band we have fitted the radial profiles with a \citep{Sersic1968} fit to determine the r.m.s of our profile data points in relation to the S\'ersic fit. The S\'ersic function is a three parameter law that is known to model very accurately the surface brightness profile of elliptical galaxies. For instance, it is known that the surface brightness profile of M87 was reported to follow a $R^{1/4}$ law \citep{deVaucouleurs1978}.
Hence, we have fitted the S\'ersic model to our data and estimated its scatter around the fit for each band\footnote{It is worth noting that our intention  with these S\'ersic fitting is not obtaining an estimation of the structural parameters of M87 in different bands. The spatial range of our observations will not allow us to do that properly. Our goal is to have an independent photometric way of measuring the level of oscillation of our photometric data in relation to a smooth surface brightness distribution.}. 
This scatter or deviation of the S\'ersic profile should be a good indication of the real photometric error in each band. The residuals to the S\'ersic fits are shown in Appendix \ref{app:sersic}. Therefore, the total errors, a combination of both zeropoint error and the photometric oscillation, are given in Table \ref{tablaerrors}.

For consistency, we also estimate the photometric errors for the wide-field data using the same technique. It is worth noting that for 2MASS, the nominal errors are bigger than our estimates. Therefore, to be conservative, we used the nominal errors provided by 2MASS photometry.

\begin{table*}
 \begin{center}
  \begin{tabular}{cccccccc} \\

 F25SFR2 & F25QTZ & F$336$W  & F$410$M &F$467$M & F$475$W &F$555$W & F$547$M  \\ 
\multicolumn{8}{c}{\cellcolor[gray]{0.8}[mag/arcsec$^2$]} \\ \hline
 $0.054$  &  $0.299$ & $0.033$  & $0.026$ & $0.030$ & $0.026$ &   $0.025$ & $0.039$  \\ 
  F$606$W & F$658N$ & F$702$W & F$814$W &  F$850$LP & $J$ & F$160$W & $K_s$\\ 
\multicolumn{8}{c}{\cellcolor[gray]{0.8}[mag/arcsec$^2$]} \\ \hline
 $0.025$ & $0.051$ &  $0.026$ & $0.026$ & $0.026$ & $0.045$ & $0.052$ & $0.059$ \\ 

 \multicolumn{2}{c}{FUV GALEX} & \multicolumn{2}{c}{NUV GALEX} & \multicolumn{2}{c}{$u$} & \multicolumn{2}{c}{$g$}   \\
 \multicolumn{8}{c}{\cellcolor[gray]{0.8}[mag/arcsec$^2$]} \\ \hline
   \multicolumn{2}{c}{$0.107	$} &  \multicolumn{2}{c}{$0.103$}  & \multicolumn{2}{c}{$0.029$} & \multicolumn{2}{c}{$0.013$} \\
    $r$ & $i$ & $z$ & $J$ & $H$ & $K_s$   \\
   \multicolumn{8}{c}{\cellcolor[gray]{0.8}[mag/arcsec$^2$]} \\ \hline
    $0.013$ &$0.014$  &  $0.018$  &  $0.026$*  &  $0.031$*  &  $0.027$* \\
  \end{tabular}\caption{Estimated photometric errors in each band. The errors in $J$, $H$ and $K_s$ (marked with asterisks) are smaller than the nominal errors provided by the 2MASS photometry. In those cases, we used the most conservative values provided by 2MASS.}\label{tablaerrors}
 \end{center}
\end{table*}

In the following sections, the properties of the stellar populations of the central region of M$87$ are analysed in two ways. First, we explore the colour gradients, focusing in the new NIR imaging. Second, we fit the SEDs using SSPs models to derive the age and metallicity of the galaxy.

\section{The inner $\sim8$ kpc of M$87$}\label{secageandmet}

We recall that the aim of this paper is to characterize the stellar population of the inner $8$ kpc of M$87$ using unprecedent HR SED data for its most inner regions from UV to IR. Accesing a large wavelength range yields to more reliable estimates of age and metallicity, even for older ages \citep{Anders2004}. A first inspection of the radial profiles in all bands shows a smooth behaviour. Consequently, no drastic changes are expected in the stellar population properties but a mild increase or decrease variation. We have checked visually that the presence of a disk (R$\sim 1\arcsec$) and extended filaments (reaching $\sim10\arcsec$) of ionized gas in the H$\alpha$ + N[II] (F658N) image \citep[][]{Ford1994,Pogge2000} do not modify the profiles of the adjacent bands, e.g. F$606$W, F$702$W (see Figure \ref{figprof}).

\subsection{Colour gradients}\label{colours}
Radial colour gradients can provide valuable constraints in the formation processes of ellipticals. While the NIR wavelength range depends mainly on the red giant-branch, the optical regime depends on stars near the main sequence turn-off point being both metallicity and age sensitive. By using the combination of optical-IR colours, we have therefore a better chance to break the age-metallicity degeneracy inherent in optical colours alone \citep[e.g.][]{Kissler-Patig2002}. Optical-IR colours together with optical colours can separate the effects of metallicity and age \citep[e.g.][]{Peletier1990}. 

\begin{figure}
 \begin{center}
   \includegraphics[scale=0.6]{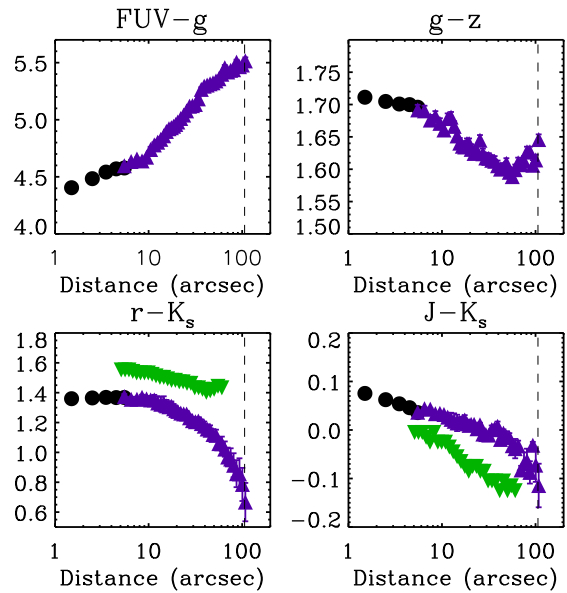} 
    \caption{Radial UV-optical, optical, optical-IR and IR colours. Black dots refer to HR photometry. Purple triangles are SDSS+2MASS photometry and green inverted triangles correspond to the radial profiles of M$87$ from \citet{Peletier1989}. The vertical dashed line marks the position of the $R_e$. \citet{Peletier1989} $J-K_s$ is shifted $-0.05$ mag for visualization.}\label{figcolour}
 \end{center}
\end{figure}

The four panels in Figure \ref{figcolour} show IR, optical and combinations of optical-IR and UV-optical colours. $r-K_s$, $g-z$ and $FUV-g$ correspond to $F606W-K_s$, $F475W-F850LP$ and $F25SFR2-F475W$ for the HST imaging of the central parts of the galaxy. The ACS filter F$475$W (F850LP) is analogous to SDSS $g$ ($z$), so the comparison with the outer region of the galaxy is straightforward. A vertical shift to the $r-K_s$, $FUV-g$ and $J-K_s$ has to be applied to account for the difference among the filters: F$606$W and $r$, NaCo $J$ and 2MASS $J$ and STIS F25SFR2 and GALEX FUV. We have overplotted the data from \citet{Peletier1989} for comparison. The observed colours agree with a mild decrease on the properties of the stellar population of the galaxy down to $R_e$. The decrement observed in all the colour profiles, except the UV colour, is in agreement with the broadband optical and NIR colours seen in \citet{Peletier1989}, although in 2MASS $J-K_s$ (purple triangles) the decrease seems less pronounced 
in our data. 
Regarding the UV colour, the FUV depends mainly on hot HB stars while the optical is dominated by main sequence and giant stars \citep{Dorman1995}. Thus, $FUV-g$ is a measure of the UV excess and, consequently, of the fraction of the stellar population that appears in the form of hot HB stars. It is seen that $FUV-g$ is bluer in the central parts of the galaxy, thus the UV excess is more intense, as expected in \citet{Dorman1995}, and then becomes more and more faint with radius. \citet{Ohl1998} found that the $FUV-B$ colour profile is flat out to $20\arcsec$ and then the colour becomes redder with increasing radial distance. The contribution of the nucleus and jet is ruled out, as they only contribute about $17\%$ of the light within $20\arcsec$. This result is in agreement with our $FUV-g$ colour.

In the inner $\sim10\arcsec$ ($\lesssim1$ kpc) the colour profiles present an almost flat slope compared to the wide-field imaging, except in the case of $J-K_s$, and only at larger distances the colours begin to drop. Further evidences of flat IR colour profiles in the inner regions of M87 down to $5\arcsec$ are found in \citet{Corbin2002}.
It is worth noting that the $g-z$ colour shows an abrupt increase with radius for R$>50\arcsec$ not observed in any other colours  in Figure \ref{figcolour}, probably due to an incorrect sky background estimation or PSF issues \citep[e.g. ][]{Tal2011}. This is also observed in the $g-r$ colour. Regarding $J-K_s$, the higher errors and dispersion are produced by the low exposure times of 2MASS data. To estimate at which radius the 2MASS photometry is realiable, we have compared our $K_s$-band profile with the deeper K-band profile kindly provided by R. L\"asker (priv. comm., \citealt{Lasker2013}). This comparison shows that both profiles are compatible in the central parts but at distances farther than $50\arcsec$ the 2MASS $K_s$ profile becomes clearly underestimated. 
The increase in $g-z$ and the departure of the shape of 2MASS profiles beyond M87's $R_e$ forced us to set the limit of our analysis to $\sim100\arcsec$. In any case, the conclusions of this work do not rely in our observations for R$>50\arcsec$ (at those distances, we have used other data from the literature).

The colour gradients seem to correlate well among them. This indicates that both optical and optical-IR radial colour gradients are likely caused by an identical physical process. As the optical-IR traces metallicity, we can therefore expect that changes in these colours indicate changes in metallicity. Consequently, the metallicity gradient is very likely the main cause for the colour gradients in the other bands as well. On the following sections we expand on this possibility.

\subsection{Age and metallicity gradients}\label{SSPfitting}
Taking advantage on the large wavelength range of our dataset, we fitted single stellar population (SSP) models to explore whether it is possible to detect stellar population gradients in the inner $8$ kpc of M$87$. SSP models are the most simple description we can make of our data. Note that this could be a rough assumption at large radii because we are considering a two-phase formation scenario (see the Introduction) and at large radii we expect the accretion of a variety of low-mass galaxies. We use the same approach as in Montes et al. (2013, submitted) to determine age and metallicity of the inner $3$ square kpc M87 globular cluster population. We describe our approach in the following subsections. 

\subsubsection{Methodology}\label{modelfit}

A SSP is defined as a single generation of coeval stars characterized by the same parameters, the most relevant being the metallicity, age and initial mass function (IMF). The observed SEDs are compared with some model SSPs to obtain information of the stellar populations of M$87$.
Since our data are integrated luminosities, we convolved the theoretical SED spectra with the filter response of our photometric filters to retrieve synthetic photometry for comparison. The computed magnitude, in the AB system, for the $i^{th}$ filter is

\begin{equation}
 m^{i}_{AB}=-2.5\log\frac{\int_{\nu}F_{\nu}\phi^{i}(\nu)d\nu}{\int_{\nu}\phi^{i}(\nu)d\nu}+8.906,
\end{equation}

where $F_{\nu}$ is the theoretical SSP SED expressed in Jy, which is a function of metallicity, age and luminosity, and $\phi^{i}$ is the response curve of the $i^{th}$ filter. Given the wide wavelength coverage of M87 to find the most suitable SSP model, a reduced-$\chi^2$ minimization approach is applied.

\begin{equation}
   \tilde{\chi}^2=\frac{1}{N-n-1}\sum_{i=1}^{15}\frac{(m_{obs,i}-m_{mod,i})^2}{\sigma_i^2},
\end{equation}

where $m_{mod,i}$ depend on age, metallicity and luminosity, $N$ is the number of photometric filters, $n$ is the number of fitted parameters and $\sigma_i$ are the observational errors in the photometry of each band. As the parameters to fit are $3$: age, metallicity and luminosity, the number of spectral bands required for the fit are at least $5$.

To illustrate the reliability of our age and metallicity estimations, we followed the standard procedure to estimate the $68\%$ and $95\%$ confidence regions. For each solution we have calculated the value of the $\tilde{\chi}^2$ for each of our models (given by Equation 2). After finding the minimum and taking into account the degrees of freedom ($N-n-1$), we have searched for all the solutions in the age and metallicity maps which are at a given distance of the minimum $\tilde{\chi}^2$ reduced value, corresponding to the $68\%$ and $95\%$ of probability as given by a chi-squared distribution table. The areas dictated by these limits represent our confidence interval solutions. Note that the degrees of freedom ($N-n-1$) vary with radius.

\begin{figure}
 \begin{center}
   \includegraphics[height=0.45\textheight]{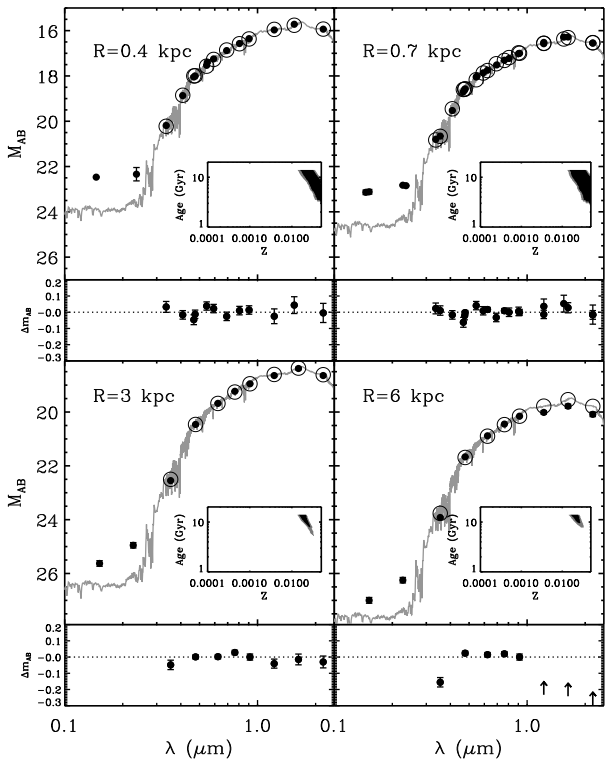} 
    \caption{SEDs at four different radii (black dots) and the best fitting BC03 models (grey lines). The inset represents the age-metallicity parameter space and the contours of the $68\%$ (black), and the $95\%$ (grey). Note the age-metallicity degeneracy. The open circles represent the convolution model with the filter response in each bandpass. At large radii, the 2MASS IR data is underestimated due to incorrect background estimation. In the lower panels, the residuals between the model and observation are shown for each SED. For R$\sim6$ kpc, the lower limits in the IR photometry show the effect of the oversubstraction of the sky.}\label{figfit}
 \end{center}
\end{figure}

The models used here are \citet[][hereafter BC03]{Bruzual2003} models. The BC03 SSP models contain $221$ spectra describing the spectral evolution of SSPs from $0.1$ Myr to $20$ Gyr for $6$ different metallicities: Z=$0.0001$, $0.0004$, $0.004$, $0.008$, $0.02$ $(Z_{\odot})$, $0.05$ for two possible IMFs, Chabrier \citep{Chabrier2003} and Salpeter \citep{Salpeter1955}. These models cover a range of wavelengths from $91$ \AA{} to $160$ $\mu m$. As the age-metallicity grid is irregular, the metallicity vector was rebinned. The grid was expanded with $200$ metallicities linearly interpolating the original SSPs. 
The maximum age of the BC03 is $20$ Gyr, in excess with the current estimate of the age of the universe $13.8$ Gyr \citep[for a flat universe with $H_0 = 67.3$ km/s/Mpc, $\Omega_M = 0.315$, $\Omega_{\Lambda} = 0.685$,][]{PlanckCollaboration2013}. Therefore, we restrict the available ages of the models to the maximum age of $\sim14$ Gyr. We use the Salpeter IMF unless indicated otherwise. BC03 models provide Chabrier and Salpeter IMFs, but using a Salpeter IMF seems a reasonable choice taking into account the recent studies favouring this kind of IMF for the most massive galaxies in the universe \citep[e.g.][]{Cenarro2003, vanDokkum2010, Ferreras2013}. 
The treatment of the TP-AGB stars in the new models (Charlot \& Bruzual 2007, priv. comm.) tend to overestimate the contribution of these stars in the NIR \citep[e.g.][]{Zibetti2013}. Nevertheless, we have confirmed that the same results presented below are obtained using the Charlot \& Bruzual (2007, priv. comm.) models. This is discussed in more detail in Appendix \ref{appendixb}.

M$87$ is known to have a UV upturn for both globular clusters \citep{Sohn2006} and the galaxy itself \citep{GildePaz2007}. This UV upturn is thought to be consequence of the presence of low-mass helium-burning stars, the so-called extreme horizontal branch \citep{Oconnell1999}. The SSPs of BC03 do not include these hot horizontal branch stars, therefore it is not expected an agreement of the models with the observations in this particular wavelength range. For this reason, we exclude the UV data points from our fits.

In Figure \ref{figfit}, the best model fits are shown for $4$ SEDs at different radial distances: $0.5$ kpc, $1$ kpc ($R_e/8$), $4$ kpc and $8$ kpc ($R_e$). The insets show the contours of the $68\%$ and the $95\%$ of probability. We can see the shift of the contours towards low metallicity as radial distance increases. For the fit at R$\sim6$ kpc, it is seen the effect of the underestimation of the $u$ and 2MASS bands (lower limits).

\subsubsection{Age and metallicity radial profiles}

 \begin{figure*}
  \begin{center}
   \includegraphics[height=0.3\textheight]{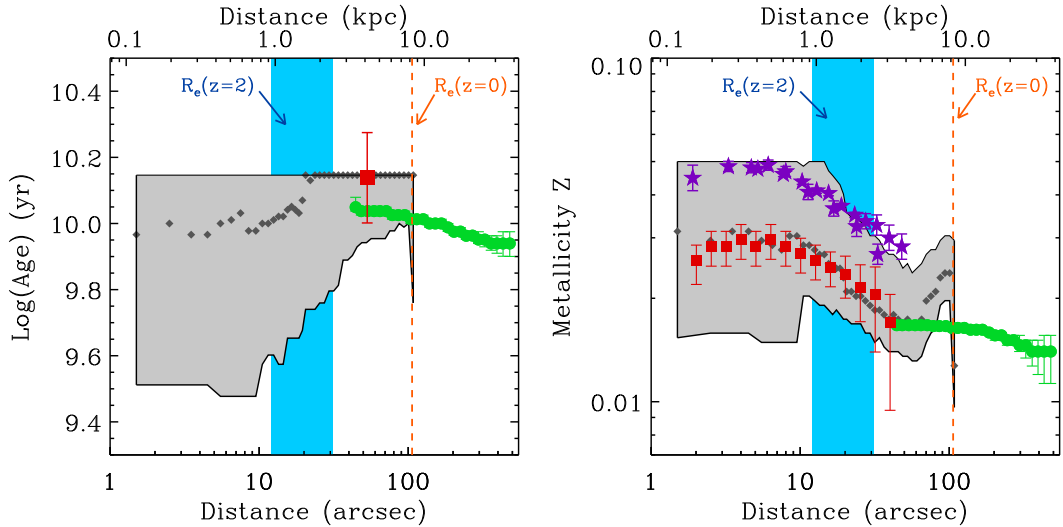} 
  \caption{Gradients of age and metallicity with radial distance for M87. The grey polygon is the $68\%$ confidence region resulting from the BC03 SSP model fit to our data and the dark grey diamonds represent the model with the minimum $\tilde{\chi}^2$. Red squares represent the data obtained with SAURON for the central region of M$87$ \citep{Kuntschner2010}. Green circles show the values of age and metallicity derived by \citet{Liu2005}. Purple stars are data from \citet{Davies1993} using absorption-line strength of $Mg_2$. We have shifted the age estimates of \citet{Kuntschner2010} from $17.7$ Gyr to $14$ Gyr. The orange dashed line indicates the position of M87's $R_e$. The blue polygon defines the range of possible values for the size of M87 at z$=2$.}\label{ageandmet}
 \end{center}
\end{figure*}

In Figure \ref{ageandmet}, the age and metallicity radial profiles of M$87$ obtained by fitting the models are shown. The grey polygon represent the $68\%$ confidence region obtained from the fits to the radial SEDs (see Section \ref{modelfit}). To illustrate the quality of the fits over the radial range, we show the minimum $\tilde{\chi}^2$ as a function of radius in Figure \ref{app:minchi} in Appendix \ref{appendixb}. In Figure \ref{ageandmet}, we have overplotted other data obtained from the literature. Purple stars are data from \citet{Davies1993} from absorption-line strength of Mg$_2$ transformed to $Z$ using the relationship provided in \citet{Casuso1996}. Red squares represent the data obtained with SAURON for the central region of M$87$ \citep{Kuntschner2010}. The red square in the age (left plot) represents the age profile derived from the SAURON data, which does not vary down to $R_e$ and, for simplicity, is represented as a single point at R$=50\arcsec$. Information at higher radial distances, obtained from the literature, was added to have an overall picture of the whole galaxy. Green circles show the values of age and metallicity derived by \citet{Liu2005}. The literature data are taken from several sources and different stellar population models or relationships (\textsc{Pegase} models from \citealt{Fioc1997, Fioc1999} for \citealt{Liu2005}; \citealt{Schiavon2007} models for \citealt{Kuntschner2010}; \citealt{Casuso1996} relationship for \citealt{Davies1993}; BC03 for this work). As the comparison among these different models is not straightforward, we do not consider a direct quantitative comparison in this work. In any case, trends and gradients can be compared qualitatively regardless of the absolute value. 

Although the optical+NIR photometry allows us to reasonably disentangle age and metallicity (see \citealt{Anders2004}), we are not able to constrain both metallicity and [$\alpha$/Fe] separately, and our metallicity profiles can indeed be affected by radial changes of [$\alpha$/Fe]. However, it is worth noting that in \citet{Kuntschner2010} the values of [$\alpha$/Fe] are $0.41\pm 0.05$ and $0.44\pm0.05$ for apertures of radii $R_e/8$ and $R_e$, respectively, indicating almost no change within errors and supporting our interpretation that the results reflect a [Fe/H] gradient. 

As seen in Figure \ref{ageandmet}, the age radial profile is compatible with no age variation down to M87's $R_e$ ($\sim8$ kpc), although in the inner $20\arcsec$ the area of the polygon is wide enough to contain ages down to $3$ Gyr due to age-metallicity degeneracy. \citet{Kuntschner2010} data (represented by a red square, shifted to $14$ Gyr) also agree with no age variation for M$87$ at these radii. Despite the large errors in our results, we can rule out young ages for large radii (where our photometrical errors are smaller) as the polygon tends to get narrower outwards, due to the smaller errors of SDSS photometry.

The metallicity profile presents a \textit{plateau} for the inner $1$ kpc, also observed in the flattening of the colour profiles (see Section \ref{colours}). Still, super solar metallicity values are expected for the central $1$ kpc while towards larger radii the metallicity seems to decrease, i.e., metallicity descend below solar abundance. \citet{Davies1993} studied out to $50\arcsec$ from the centre of M$87$ using absorption-line strengths in spectra (purple stars). Although our \textit{plateau} could be argued to be a consequence of reaching the maximum metallicity provided in BC03 (see below), it is also observed in their data, as well as the decrease of the metallicity with radius. This strongly suggests that the inner plateau in the metallicity is real.

\citet{Kuntschner2010} using SAURON measured the properties of M$87$ using integral field spectroscopy (red squares). Their metallicity profile surprisingly resembles our profile too, although \citet{Kuntschner2010} absolute values in the central parts are lower. It also reproduces the \textit{plateau} at R$<10\arcsec$. Their estimated ages ($\sim17$ Gyr) are rescaled according to the maximum age set in Section \ref{modelfit}: $14$ Gyr. Note that the $H_\beta$ line-strengths are significantly and systematically weaker than the stellar population model predictions \citep[see][]{Kuntschner2010} causing that the derived ages to saturate at the maximum allowed age of the models ($\sim17$ Gyr). Therefore we have to be cautious deriving conclusions using this profile. Assuming that no changes are made by the indetermination of ages, we can say there is an agreement between the metallicity profiles.

The gradient of the entire profile, calculated as a linear fit of the metallicity [Z/H] vs. $\log \rm{R}$ is $-0.26\pm0.10$ ($-0.25\pm 0.09$ for a Chabrier IMF). The Mg$_2$ data obtained by \citet{Davies1993} also show a gradient in metallicity compatible with our results ($-0.21\pm0.01$). They also find a slightly positive slope for the $H_{\beta}$ profile.

Furthermore, we have also explored the metallicity profile imposing an age constraint, from $10$ to $14$ Gyr. It seems reasonable to assume old ages for the bulk of stars of the galaxy as seen in previous studies \citep[e.g.][]{Liu2005,Kuntschner2010}. The results are presented in Figure \ref{app:ageandmet}. The main difference compared to Figure \ref{ageandmet} is that the width of the allowed metallicity solutions has decreased in the range from $10\arcsec$ to $20\arcsec$. Consequently, this result supports the idea that the \textit{plateau} is a real characteristic of the metallicity profile of M87 and not an artifact due to limitations of the models. In this case, the gradient is $-0.27\pm0.08$.

Our results are complemented with photometric derived ages and metallicities from \citet{Liu2005} up to $5\,R_e$. Their ages are in agreement with our results in the overlapping region, as well as with our metallicities. The gradient in metallicity of \citet{Liu2005} is $-0.080 \pm 0.006$, shallower than what it is found here. However, there is a hint in our data of a change in the slope of the profile towards $60$ arcsec ($\sim 5$ kpc) to a much shallower gradient, as observerd in \citet{Liu2005}. This change in the slope may be originated by the evolution of the galaxy as we will discuss below. However, note that this change in our metallicity profile coincides with the change of behaviour in the $g-z$ colour (while the increase in the minimum $\tilde{\chi}^2$ in Figure \ref{app:minchi} is due to the underestimation of the $u$ and IR photometry), therefore the conclusions of our work are based in \citet{Liu2005} results.  

\section{Discussion}\label{discussion}

Using HR photometry combined with wide field literature data, we have derived ages and metallicities for the inner $R_e$ of M$87$ using SED fitting. We found that changes in metallicity can explain why colours become bluer towards the outer parts of the galaxy. The metallicity profile decreases with radius while ages are constant down to M87's $R_e$. The metallicity radial profile of M87 presents three different zones: in the innermost kpc a \textit{plateau} with supersolar metallicity, a decline in metallicity down to $\sim 5$ kpc which is followed by an almost flat region with near solar metallicity. The shape of the metallicity profile is independent of the IMF used, Salpeter or Chabrier, as shown in Appendix \ref{appendixb}. Down to the $R_e$ of M87, the age of the stellar population is old ($>10$ Gyr) and the lack of an age gradient is compatible with a rapid episode of star formation in the past.

\subsection{Comparison with previous studies}

As mentioned before, our age and metallicity estimates are in agreement with previous determinations of these values by \citet{Kuntschner2010} and \citet{Liu2005}. In addition, studies of globular clusters in M$87$ \citep[e.g.][]{Cohen1998, Jordan2002} and their link to the galaxy itself \citep{Kundu1999, Harris2009} also support old ages for the bulk of stars of the galaxy, similar to what we find in this work. Our metallicities are supersolar and show evidence for a gradient towards the outer parts of the galaxy as expected from its colour profiles (Section \ref{colours}, and \citealt{Peletier1990}). This is also observed in \citet{Davies1993, Liu2005} and in \citet{Kuntschner2010}.

The age of the stellar population of M87 shows no variation in agreement with the lack of gradient found in massive ellipticals in, for example, \citet{Sanchez-Blazquez2006, Sanchez-Blazquez2007, Spolaor2008, Spolaor2010, Kuntschner2010, Loubser2012}. It is worth noting however that, \citet{Davies1993} find a slightly positive H$_{\beta}$ gradient, but possibly related to contamination from H$_{\beta}$ emission observed in this galaxy. 
Conversely, \citet{Liu2005} found a mild decrease in age for the outer parts of the galaxy (R$\gtrsim R_e/2$) compatible with some galaxies in \citet{Greene2012} but contrary to the results found in \citet{Sanchez-Blazquez2007,LaBarbera2010,LaBarbera2012} where null or positive age gradients are observed in early-type galaxies.

Our global metallicity gradient is $-0.26\pm0.10$. Similarly, \citet{Davies1993} found a metallicity gradient of $-0.21 \pm 0.01$ using absorption-line measurements of Mg$_2$ of the central part of M$87$ up to $\sim 50\arcsec$. As we show now, gradients obtained in other galaxies are similar. The mean metallicity gradient of \citet{Kuntschner2010} sample of old galaxies ($>8$ Gyr) is $-0.25\pm0.11$. For \citet{Loubser2012} sample of bright cluster galaxies, the metallicity gradient is $-0.29\pm0.06$ and for galaxies in the Coma cluster of \citet{Sanchez-Blazquez2006} is $-0.33\pm0.16$. However, for R$>R_e$ the gradient turns into $-0.080\pm 0.006$ for M87 \citep{Liu2005}, compatible with the gradients of low-mass early-type galaxies \citep{Spolaor2009}. 
 
In relation to the different gradients observed in the metallicity profile, NGC~$4889$, one of the two central bright cluster galaxies of the Coma cluster show comparable gradients to M87's \citep{Coccato2010}. Two different metallicity gradients are observed for the inner part (R$\lesssim1.2R_e$; $-0.35\pm0.02$) and outer part (R$\gtrsim1.2R_e$; $-0.1\pm0.2$). Moreover, \citet{Baes2007} find a also a break in the slopes of the metallicity gradients, getting shallower at larger radii. A similar change in gradient for a radius $\sim R_e$ is observed in our results. This potentially indicates that different mechanisms formed the central parts of the galaxy and the outer regions.
 
Summarizing, the results found in our work are similar to those found by other authors in M87 and in other massive early-type galaxies. It is worth noting that the results for M87 might be generalized only to the case of galaxies being the central objects in dark matter halos, and not massive galaxies as a whole, as central galaxies are observed to have different properties than satellites \citep[e.g.][]{Pasquali2010}.

\subsection{How can M87's age and metallicity profiles help us to understand its formation?}

According to the current scenario of galaxy formation and evolution, massive galaxies were assembled in two different phases \citep[e.g.][]{Naab2009}. In the first phase, the innermost regions formed the majority of their stars at high redshift and on short timescales \citep[e.g.][]{Thomas2005}. During this dissipative \textquotedblleft monolithic collapse\textquotedblright, stars were efficiently created while the gas sank to the  centre of the forming galaxy. As a consequence of this process, a negative radial metallicity gradient was created with the central stars being more metal-rich than the stars in the outskirts. For example, \citet{Larson1974} estimated a metallicity gradient of $\sim -0.35$, $\sim -0.5$ for \citet{Carlberg1984} and $-1.0$ in \citet{Kobayashi2004}. However, the rapid formation of the galaxy produced almost no noticeable variation in age.

In the second phase of the galaxy formation, there is a size growth of the galaxies driven by the accretion of low-mass satellites. These satellites, mostly depleted of gas, end located in the outskirts of these galaxies \citep[e.g.][]{Khochfar2006}. For this reason, one would expect a significant change in the stellar properties with radius as the stars of these merged satellites are mainly located in the periphery of the main galaxy due to their higher angular momentum. These stars should be relatively evolved and no star formation is expected to occur if mergers are dry. Consequently, the new accreted stars are not expected to produce significant age variations in the galaxy. As the stars of these low-mass satellites will be metal-poor (as indicated by observations, e.g. \citealt{Spolaor2009}), the radial metallicity gradient is predicted to change due to the contribution of metal-poor stars from the shreded satellites \citep{Naab2009}. For instance, one could expect a steepening of the metallicity 
gradient in the outskirts. In other words, we expect a steep negative radial gradient in the inner region of the galaxy while in the outer parts a change of gradient is suggested. On what follows, we discuss how our results fit in this picture. 

One of the most intriguing results that we have found in this paper is the presence of a \textit{plateau} in the colour profiles and in the metallicity gradient of M87 at R$\lesssim1$ kpc. This flattening in the metallicity profile is not expected in the models of a dissipative \textquotedblleft monolithic collapse\textquotedblright. Is this \textit{plateau} a real property of the metallicity profile? As shown in Figure \ref{ageandmet}, \citet{Davies1993} and \citet{Kuntschner2010} found the same behaviour using ground-based data affected by seeing up to $3\arcsec$ and $1\arcsec$, respectively. Our HR results, show that the metallicity flattening is not caused by the blurring effect of their seeing. Intriguingly, this inner \textit{plateau} is not a unique feature of M87. In fact, this kind of behaviour is also observed in \citet{Sanchez-Blazquez2007}, where a similar flattening for the inner $\sim9\arcsec$ ($\sim 2.5$ kpc) is present in the metallicity profile of the field massive galaxy NGC~1600.

How can we explain this metallicity \textit{plateau} ($\sim 1$ kpc) in the context of massive galaxy formation? The timescales for dissipative collapse are of $<0.5-1$ Gyr \citep{Pipino2008}. To explore the feasibility of these short timescales for the formation of the innermost region of M87, we have estimated the mass enclosed by the \textit{plateau} using Equation $15$ in \citet{Trujillo2001}. Almost $10\%$ of the mass of M87 is within $\sim1$ kpc. That is $\sim7\times10^{10}\,M_{\odot}$, assuming that the total stellar mass of M87 is $6.8\,\pm\,1.1\times10^{11}\,M_{\odot}$ \citep[taken from][]{Forte2012}. So, the star formation rate within $\sim1$ kpc should have been $\gtrsim 70-140\,\rm{M}_{\odot}\,\rm{yr}^{-1}$. Star formation rates as high as $\sim3000\,\rm{M}_{\odot}\,\rm{yr}^{-1}$ have been found in a massive starbursts galaxies at $z\sim6$ \citep[][]{Riechers2013}. Consequently, a rapid monolithic collapse looks like a very plausible scenario. New evidence in the recently discovered relic galaxy NGC~1277 \citep{Trujillo2013} also support this picture.

Moreover, the size of this \textit{plateau} can provide clues about the size evolution of M87. It is widely known that high-redshift massive galaxies were smaller in size than their present-day massive counterparts \citep[e.g.][]{Daddi2005, Trujillo2006, Buitrago2008}. Possibly, this evolution in size has left a signature on the properties of the stellar populations of M87. For this reason, we have calculated the $R_e$ that a galaxy like M87 would have had at $z=2$. Using the following equation for size evolution \citep[in][]{Trujillo2011}:

\begin{equation}
 \log R_f= \log R_i +\Delta \log R |_{M_{i,fixed}}+ C \log(M_f/M_i)
\end{equation}

where $R_f$ and $M_f$ are the radius and the mass at $z=0$, and $R_i$ and $M_i$ the radius and the mass at a given $z$. $\Delta \log R |_{M_{i,fixed}}$ accounts for the variation of the radius at a fixed mass, while the term $C \log(M_f/M_i)$ reflects the change in radius due to the variation of the mass as a consequence of the accretion of satellite galaxies. Many of the previous parameters are fixed by the observations. For instance, we know that galaxies have doubled their mass since $z=2$ \citep{vanDokkum2010a}, $C=0.56$ from \citet{Shen2003} and the evolution in radius at a fixed stellar mass is described by $R_i=R_f(1+z)^{\beta}$. The $\beta$ parameter ranges from $-0.75$ to $-1.62$ \citep[see][and references therein]{Oser2012}. We assume that $R_f$ is the observed $R_e$ of M87 at $z=0$ ($106\farcs2$, \citealt{Falcon-Barroso2011}). The above values provide us with a region of potential values for the size ($R_e$) of M87 at $z=2$. We obtain that M87 $R_e$ at $z=2$ should have been between $0.94$ kpc to 
$2.5$ kpc in agreement with the position of the turning point of the \textit{plateau} at $\sim1$ kpc, given that the formation of this flattening was occured very early-on. This region is shaded in light blue in Figure \ref{ageandmet}. As discussed in \citet{Falcon-Barroso2011}, the value of the $R_e$ of M87 is uncertain. According to the size distribution of galaxies in \citet{Shen2003}, we expect a higher $R_e$ ($\sim10$ kpc). Therefore, the values obtained for the size of M87 at z$\sim2$ could range between $1$ kpc and $3$ kpc, still compatible with our results. This matches the picture that this flattening of the metallicity provides a signature of the size of the proto-galaxy at $z=2$, after its quick formation.

Concerning the rest of the galaxy, beyond the inner \textit{plateau} there is a zone of declining metallicity observed from $\sim1$ kpc to $\sim5$ kpc (see Figure \ref{ageandmet}). Although the global gradient is $-0.26\pm0.10$, the local gradient for this region, between $1$ to $5$ kpc, is steeper:$-0.4\pm 0.2$. This steep metallicity gradient down to $R_e$ is reminiscent of the predictions of the dissipative quasi-monolithic collapse mentioned before \citep[e.g.][]{Larson1974, Carlberg1984, Kobayashi2004, Pipino2008}. The slopes range between $\sim -1$ to $\sim -0.35$ for different models, in agreement with the gradient found in M87. While this dissipative process tend to create metallicity gradients, mergers between galaxies destroy these gradients \citep[e.g.][]{White1980, Kobayashi2004}. Our derived metallicity gradient is compatible with the $\sim -0.3$ for galaxies built in a dissipative collapse. Consequently, the observed [Z/H] gradient is probably the consequence of the dissipative formation of 
this zone of the galaxy. In short, the inner $5$ kpc of M87 seems to have formed through a dissipative collapse, with the innermost $1$ kpc having formed in a extremely short timescale as suggested by the \textit{plateau} in its metallicity gradient. 

Outwards, at $\gtrsim 5$ kpc, there is a change in the metallicity gradient that becomes flat at these distances. This evidence of a change of slope supports the scenario, as discussed above, where the centre of the galaxy formed via a rapid dissipative process whereas the rest of the galaxy was built via accretion of low-mass satellites. If these low-mass systems had been accreted in a more massive galaxy, different gradients are a natural outcome as the stars tend to populate the outer parts of the galaxy as seen in, for example, \citet{Coccato2010}. This flattening suggests that most of the mass that built the outskirsts of the galaxy came from satellites of similar metallicity (sub-solar). Moreover, this gradient turning point denotes where the influence of the massive proto-galaxy ends (i.e. at $2-3\,R_e$ at $z=2$).

This behaviour of the metallicity gradient is consistent with recent studies \citep[e.g.][]{Daddi2005, Trujillo2006, Trujillo2011} where early-type galaxies have sizes a factor of $4$ larger than their counterparts at $z\sim2$. The central stellar mass density of the massive galaxies at high-$z$ do not differ much from the central density of nearby galaxies \citep{Hopkins2009}. Consequently the majority of the evolution has taken place in their outer regions. The high dispersion in gradients for high-mass galaxies observed in \citet{Spolaor2009} also points to minor mergers as a mechanism for the build-up of the galaxy. 

In summary, the metallicity profile presents three different zones that correspond to two separate mechanisms for the formation of the galaxy. The inner parts of the galaxy, R$\lesssim5$ kpc, are likely formed as a result of a dissipative event in the past. They have formed very rapidly, especially the very central $\sim1$ kpc which seems to be the consequence of a very intense burst of star formation. Oppositely, the outer galaxy, R$\gtrsim5$ kpc may be the result of the accretion of low-mass galaxies. 

\section{Summary and conclusions}\label{summary}

Using HR imaging from \emph{HST} and adaptive optics from NaCo at VLT, we derived radial SEDs for the central $\sim1$ kpc of M$87$. This dataset was complemented with wide field imaging from GALEX, SDSS and 2MASS to reach up to M87's $R_e$ ($\sim8$ kpc). Constructing colours from this dataset, radial changes in the stellar populations are seen. Studying this in more detail through SED fitting, we found that radial variations trace changes in metallicity (Section \ref{SSPfitting}). Adding data from the literature up to $5R_e$, three different behaviours for the metallicity profile are observed: a central flattening or \textit{plateau} at R$\lesssim1$ kpc, a decline in metallicity down to $R_e$ and a change in the metallicity gradient for R$>R_e$. The central metallicity gradient agrees with previous estimates of M$87$ with absorption line strengths. The metallicity structure of M87 is consistent with the current scenario of galaxy formation where the central part of the galaxy had formed first and the outer 
part was accreted via dry minor merging.

\section*{Acknowledgements}
We thank the anonymous referee for his/her detailed and constructive comments that improved the paper. We also thank A. Vazdekis and P. S\'{a}nchez-Bl\'{a}zquez for many useful comments on stellar population gradients, H. Kuntschner for kindly providing the SAURON metallicity profile of M87 and A. Gil de Paz, C.-W. Chen and R. L\"{a}sker for providing the GALEX, SDSS and K radial profiles of M87, respectively. This work is partially funded by the Spanish MEC project AYA2007-60235 and AYA2010-21322-C03-02.

\bibliography{m87gal}

\begin{thebibliography}{86}
\expandafter\ifx\csname natexlab\endcsname\relax\def\natexlab#1{#1}\fi

\bibitem[{{Anders} {et~al}\mbox{.}(2004){Anders}, {Bissantz},
  {Fritze-v.~Alvensleben}, \& {de Grijs}}]{Anders2004}
{Anders} P., {Bissantz} N., {Fritze-v.~Alvensleben} U., {de Grijs} R., 2004,
  \mnras, 347, 196

\bibitem[{{Baes} {et~al}\mbox{.}(2010){Baes}, {Clemens}, {Xilouris}, {Fritz},
  {Cotton}, {Davies}, {Bendo}, {Bianchi}, {Cortese}, {de Looze}, {Pohlen},
  {Verstappen}, {B{\"o}hringer}, {Bomans}, {Boselli}, {Corbelli}, {Dariush},
  {di Serego Alighieri}, {Fadda}, {Garcia-Appadoo}, {Gavazzi}, {Giovanardi},
  {Grossi}, {Hughes}, {Hunt}, {Jones}, {Madden}, {Pierini}, {Sabatini},
  {Smith}, {Vlahakis}, \& {Zibetti}}]{Baes2010}
{Baes} M. {et~al.}, 2010, \aap, 518, L53

\bibitem[{{Baes} {et~al}\mbox{.}(2007){Baes}, {Sil'chenko}, {Moiseev}, \&
  {Manakova}}]{Baes2007}
{Baes} M., {Sil'chenko} O.~K., {Moiseev} A.~V., {Manakova} E.~A., 2007, \aap,
  467, 991

\bibitem[{{Bertin} \& {Arnouts}(1996)}]{Bertin1996}
{Bertin} E., {Arnouts} S., 1996, \aaps, 117, 393

\bibitem[{{Blakeslee} {et~al}\mbox{.}(2001){Blakeslee}, {Lucey}, {Barris},
  {Hudson}, \& {Tonry}}]{Blakeslee2001}
{Blakeslee} J.~P., {Lucey} J.~R., {Barris} B.~J., {Hudson} M.~J., {Tonry}
  J.~L., 2001, \mnras, 327, 1004

\bibitem[{{Bruzual} \& {Charlot}(2003)}]{Bruzual2003}
{Bruzual} G., {Charlot} S., 2003, \mnras, 344, 1000

\bibitem[{{Buitrago} {et~al}\mbox{.}(2008){Buitrago}, {Trujillo}, {Conselice},
  {Bouwens}, {Dickinson}, \& {Yan}}]{Buitrago2008}
{Buitrago} F., {Trujillo} I., {Conselice} C.~J., {Bouwens} R.~J., {Dickinson}
  M., {Yan} H., 2008, \apjl, 687, L61

\bibitem[{{Cardelli}, {Clayton} \& {Mathis}(1989){Cardelli}, {Clayton}, \&
  {Mathis}}]{Cardelli1989}
{Cardelli} J.~A., {Clayton} G.~C., {Mathis} J.~S., 1989, \apj, 345, 245

\bibitem[{{Carlberg}(1984)}]{Carlberg1984}
{Carlberg} R.~G., 1984, \apj, 286, 416

\bibitem[{{Casuso} {et~al}\mbox{.}(1996){Casuso}, {Vazdekis}, {Peletier}, \&
  {Beckman}}]{Casuso1996}
{Casuso} E., {Vazdekis} A., {Peletier} R.~F., {Beckman} J.~E., 1996, \apj, 458,
  533

\bibitem[{{Cenarro} {et~al}\mbox{.}(2003){Cenarro}, {Gorgas}, {Vazdekis},
  {Cardiel}, \& {Peletier}}]{Cenarro2003}
{Cenarro} A.~J., {Gorgas} J., {Vazdekis} A., {Cardiel} N., {Peletier} R.~F.,
  2003, \mnras, 339, L12

\bibitem[{{Chabrier}(2003)}]{Chabrier2003}
{Chabrier} G., 2003, \apjl, 586, L133

\bibitem[{{Chen} {et~al}\mbox{.}(2010){Chen}, {C{\^o}t{\'e}}, {West}, {Peng},
  \& {Ferrarese}}]{Chen2010}
{Chen} C.-W., {C{\^o}t{\'e}} P., {West} A.~A., {Peng} E.~W., {Ferrarese} L.,
  2010, \apjs, 191, 1

\bibitem[{{Chiosi} \& {Carraro}(2002)}]{Chiosi2002}
{Chiosi} C., {Carraro} G., 2002, \mnras, 335, 335

\bibitem[{{Coccato}, {Gerhard} \& {Arnaboldi}(2010){Coccato}, {Gerhard}, \&
  {Arnaboldi}}]{Coccato2010}
{Coccato} L., {Gerhard} O., {Arnaboldi} M., 2010, \mnras, 407, L26

\bibitem[{{Cohen}(1986)}]{Cohen1986}
{Cohen} J.~G., 1986, \aj, 92, 1039

\bibitem[{{Cohen}, {Blakeslee} \& {Ryzhov}(1998){Cohen}, {Blakeslee}, \&
  {Ryzhov}}]{Cohen1998}
{Cohen} J.~G., {Blakeslee} J.~P., {Ryzhov} A., 1998, \apj, 496, 808

\bibitem[{{Corbin}, {O'Neil} \& {Rieke}(2002){Corbin}, {O'Neil}, \&
  {Rieke}}]{Corbin2002}
{Corbin} M.~R., {O'Neil} E., {Rieke} M.~J., 2002, \aj, 124, 183

\bibitem[{{Daddi} {et~al}\mbox{.}(2005){Daddi}, {Renzini}, {Pirzkal},
  {Cimatti}, {Malhotra}, {Stiavelli}, {Xu}, {Pasquali}, {Rhoads}, {Brusa}, {di
  Serego Alighieri}, {Ferguson}, {Koekemoer}, {Moustakas}, {Panagia}, \&
  {Windhorst}}]{Daddi2005}
{Daddi} E. {et~al.}, 2005, \apj, 626, 680

\bibitem[{{Davies}, {Sadler} \& {Peletier}(1993){Davies}, {Sadler}, \&
  {Peletier}}]{Davies1993}
{Davies} R.~L., {Sadler} E.~M., {Peletier} R.~F., 1993, \mnras, 262, 650

\bibitem[{{de Vaucouleurs} \& {Nieto}(1978)}]{deVaucouleurs1978}
{de Vaucouleurs} G., {Nieto} J.-L., 1978, \apj, 220, 449

\bibitem[{{Dorman}, {O'Connell} \& {Rood}(1995){Dorman}, {O'Connell}, \&
  {Rood}}]{Dorman1995}
{Dorman} B., {O'Connell} R.~W., {Rood} R.~T., 1995, \apj, 442, 105

\bibitem[{{Eggen}, {Lynden-Bell} \& {Sandage}(1962){Eggen}, {Lynden-Bell}, \&
  {Sandage}}]{Eggen1962}
{Eggen} O.~J., {Lynden-Bell} D., {Sandage} A.~R., 1962, \apj, 136, 748

\bibitem[{{Falc{\'o}n-Barroso} {et~al}\mbox{.}(2011){Falc{\'o}n-Barroso}, {van
  de Ven}, {Peletier}, {Bureau}, {Jeong}, {Bacon}, {Cappellari}, {Davies}, {de
  Zeeuw}, {Emsellem}, {Krajnovi{\'c}}, {Kuntschner}, {McDermid}, {Sarzi},
  {Shapiro}, {van den Bosch}, {van der Wolk}, {Weijmans}, \&
  {Yi}}]{Falcon-Barroso2011}
{Falc{\'o}n-Barroso} J. {et~al.}, 2011, \mnras, 417, 1787

\bibitem[{{Ferreras} {et~al}\mbox{.}(2013){Ferreras}, {La Barbera}, {de la
  Rosa}, {Vazdekis}, {de Carvalho}, {Falc{\'o}n-Barroso}, \&
  {Ricciardelli}}]{Ferreras2013}
{Ferreras} I., {La Barbera} F., {de la Rosa} I.~G., {Vazdekis} A., {de
  Carvalho} R.~R., {Falc{\'o}n-Barroso} J., {Ricciardelli} E., 2013, \mnras,
  429, L15

\bibitem[{{Fioc} \& {Rocca-Volmerange}(1997)}]{Fioc1997}
{Fioc} M., {Rocca-Volmerange} B., 1997, \aap, 326, 950

\bibitem[{{Fioc} \& {Rocca-Volmerange}(1999)}]{Fioc1999}
---, 1999, ArXiv Astrophysics e-prints

\bibitem[{{Ford} {et~al}\mbox{.}(1994){Ford}, {Harms}, {Tsvetanov}, {Hartig},
  {Dressel}, {Kriss}, {Bohlin}, {Davidsen}, {Margon}, \& {Kochhar}}]{Ford1994}
{Ford} H.~C. {et~al.}, 1994, \apjl, 435, L27

\bibitem[{{Forte}, {Vega} \& {Faifer}(2012){Forte}, {Vega}, \&
  {Faifer}}]{Forte2012}
{Forte} J.~C., {Vega} E.~I., {Faifer} F., 2012, \mnras, 421, 635

\bibitem[{{Franx} \& {Illingworth}(1990)}]{Franx1990}
{Franx} M., {Illingworth} G., 1990, \apjl, 359, L41

\bibitem[{{Fruchter} \& {Hook}(2002)}]{Fruchter2002}
{Fruchter} A.~S., {Hook} R.~N., 2002, \pasp, 114, 144

\bibitem[{{Gil de Paz} {et~al}\mbox{.}(2007){Gil de Paz}, {Boissier}, {Madore},
  {Seibert}, {Joe}, {Boselli}, {Wyder}, {Thilker}, {Bianchi}, {Rey}, {Rich},
  {Barlow}, {Conrow}, {Forster}, {Friedman}, {Martin}, {Morrissey}, {Neff},
  {Schiminovich}, {Small}, {Donas}, {Heckman}, {Lee}, {Milliard}, {Szalay}, \&
  {Yi}}]{GildePaz2007}
{Gil de Paz} A. {et~al.}, 2007, \apjs, 173, 185

\bibitem[{{Greene} {et~al}\mbox{.}(2012){Greene}, {Murphy}, {Comerford},
  {Gebhardt}, \& {Adams}}]{Greene2012}
{Greene} J.~E., {Murphy} J.~D., {Comerford} J.~M., {Gebhardt} K., {Adams}
  J.~J., 2012, \apj, 750, 32

\bibitem[{{Harris}(2009)}]{Harris2009}
{Harris} W.~E., 2009, \apj, 703, 939

\bibitem[{{Hopkins} {et~al}\mbox{.}(2009){Hopkins}, {Bundy}, {Murray},
  {Quataert}, {Lauer}, \& {Ma}}]{Hopkins2009}
{Hopkins} P.~F., {Bundy} K., {Murray} N., {Quataert} E., {Lauer} T.~R., {Ma}
  C.-P., 2009, \mnras, 398, 898

\bibitem[{{Jarrett} {et~al}\mbox{.}(2003){Jarrett}, {Chester}, {Cutri},
  {Schneider}, \& {Huchra}}]{Jarrett2003}
{Jarrett} T.~H., {Chester} T., {Cutri} R., {Schneider} S.~E., {Huchra} J.~P.,
  2003, \aj, 125, 525

\bibitem[{{Jord{\'a}n} {et~al}\mbox{.}(2002){Jord{\'a}n}, {C{\^o}t{\'e}},
  {West}, \& {Marzke}}]{Jordan2002}
{Jord{\'a}n} A., {C{\^o}t{\'e}} P., {West} M.~J., {Marzke} R.~O., 2002, \apjl,
  576, L113

\bibitem[{{Khochfar} \& {Silk}(2006)}]{Khochfar2006}
{Khochfar} S., {Silk} J., 2006, \mnras, 370, 902

\bibitem[{{Kissler-Patig}, {Brodie} \& {Minniti}(2002){Kissler-Patig},
  {Brodie}, \& {Minniti}}]{Kissler-Patig2002}
{Kissler-Patig} M., {Brodie} J.~P., {Minniti} D., 2002, \aap, 391, 441

\bibitem[{{Kobayashi}(2004)}]{Kobayashi2004}
{Kobayashi} C., 2004, \mnras, 347, 740

\bibitem[{{Kundu} {et~al}\mbox{.}(1999){Kundu}, {Whitmore}, {Sparks},
  {Macchetto}, {Zepf}, \& {Ashman}}]{Kundu1999}
{Kundu} A., {Whitmore} B.~C., {Sparks} W.~B., {Macchetto} F.~D., {Zepf} S.~E.,
  {Ashman} K.~M., 1999, \apj, 513, 733

\bibitem[{{Kuntschner} {et~al}\mbox{.}(2010){Kuntschner}, {Emsellem}, {Bacon},
  {Cappellari}, {Davies}, {de Zeeuw}, {Falc{\'o}n-Barroso}, {Krajnovi{\'c}},
  {McDermid}, {Peletier}, {Sarzi}, {Shapiro}, {van den Bosch}, \& {van de
  Ven}}]{Kuntschner2010}
{Kuntschner} H. {et~al.}, 2010, \mnras, 408, 97

\bibitem[{{La Barbera} {et~al}\mbox{.}(2010){La Barbera}, {De Carvalho}, {De La
  Rosa}, {Gal}, {Swindle}, \& {Lopes}}]{LaBarbera2010}
{La Barbera} F., {De Carvalho} R.~R., {De La Rosa} I.~G., {Gal} R.~R.,
  {Swindle} R., {Lopes} P.~A.~A., 2010, \aj, 140, 1528

\bibitem[{{La Barbera} {et~al}\mbox{.}(2012){La Barbera}, {Ferreras}, {de
  Carvalho}, {Bruzual}, {Charlot}, {Pasquali}, \& {Merlin}}]{LaBarbera2012}
{La Barbera} F., {Ferreras} I., {de Carvalho} R.~R., {Bruzual} G., {Charlot}
  S., {Pasquali} A., {Merlin} E., 2012, \mnras, 426, 2300

\bibitem[{{Larson}(1974)}]{Larson1974}
{Larson} R.~B., 1974, \mnras, 166, 585

\bibitem[{{L{\"a}sker}, {Ferrarese} \& {van de Ven}(2013){L{\"a}sker},
  {Ferrarese}, \& {van de Ven}}]{Lasker2013}
{L{\"a}sker} R., {Ferrarese} L., {van de Ven} G., 2013, ArXiv
  e-prints:1311.1530

\bibitem[{{Liu} {et~al}\mbox{.}(2005){Liu}, {Zhou}, {Ma}, {Wu}, {Yang}, {Li},
  \& {Chen}}]{Liu2005}
{Liu} Y., {Zhou} X., {Ma} J., {Wu} H., {Yang} Y., {Li} J., {Chen} J., 2005,
  \aj, 129, 2628

\bibitem[{{Loubser} \& {S{\'a}nchez-Bl{\'a}zquez}(2012)}]{Loubser2012}
{Loubser} S.~I., {S{\'a}nchez-Bl{\'a}zquez} P., 2012, \mnras, 425, 841

\bibitem[{{Merlin} {et~al}\mbox{.}(2012){Merlin}, {Chiosi}, {Piovan}, {Grassi},
  {Buonomo}, \& {Barbera}}]{Merlin2012}
{Merlin} E., {Chiosi} C., {Piovan} L., {Grassi} T., {Buonomo} U., {Barbera}
  F.~L., 2012, \mnras, 427, 1530

\bibitem[{{Naab}, {Johansson} \& {Ostriker}(2009){Naab}, {Johansson}, \&
  {Ostriker}}]{Naab2009}
{Naab} T., {Johansson} P.~H., {Ostriker} J.~P., 2009, \apjl, 699, L178

\bibitem[{{O'Connell}(1999)}]{Oconnell1999}
{O'Connell} R.~W., 1999, \araa, 37, 603

\bibitem[{{Ogando} {et~al}\mbox{.}(2005){Ogando}, {Maia}, {Chiappini},
  {Pellegrini}, {Schiavon}, \& {da Costa}}]{Ogando2005}
{Ogando} R.~L.~C., {Maia} M.~A.~G., {Chiappini} C., {Pellegrini} P.~S.,
  {Schiavon} R.~P., {da Costa} L.~N., 2005, \apjl, 632, L61

\bibitem[{{Ohl} {et~al}\mbox{.}(1998){Ohl}, {O'Connell}, {Bohlin}, {Collins},
  {Dorman}, {Fanelli}, {Neff}, {Roberts}, {Smith}, \& {Stecher}}]{Ohl1998}
{Ohl} R.~G. {et~al.}, 1998, \apjl, 505, L11

\bibitem[{{Oser} {et~al}\mbox{.}(2012){Oser}, {Naab}, {Ostriker}, \&
  {Johansson}}]{Oser2012}
{Oser} L., {Naab} T., {Ostriker} J.~P., {Johansson} P.~H., 2012, \apj, 744, 63

\bibitem[{{Pasquali} {et~al}\mbox{.}(2010){Pasquali}, {Gallazzi}, {Fontanot},
  {van den Bosch}, {De Lucia}, {Mo}, \& {Yang}}]{Pasquali2010}
{Pasquali} A., {Gallazzi} A., {Fontanot} F., {van den Bosch} F.~C., {De Lucia}
  G., {Mo} H.~J., {Yang} X., 2010, \mnras, 407, 937

\bibitem[{{Peletier}(1989)}]{Peletier1989}
{Peletier} R.~F., 1989, PhD thesis, , University of Groningen, The Netherlands,
  (1989)

\bibitem[{{Peletier}, {Valentijn} \& {Jameson}(1990){Peletier}, {Valentijn}, \&
  {Jameson}}]{Peletier1990}
{Peletier} R.~F., {Valentijn} E.~A., {Jameson} R.~F., 1990, \aap, 233, 62

\bibitem[{{Pipino} {et~al}\mbox{.}(2010){Pipino}, {D'Ercole}, {Chiappini}, \&
  {Matteucci}}]{Pipino2010}
{Pipino} A., {D'Ercole} A., {Chiappini} C., {Matteucci} F., 2010, \mnras, 407,
  1347

\bibitem[{{Pipino}, {D'Ercole} \& {Matteucci}(2008){Pipino}, {D'Ercole}, \&
  {Matteucci}}]{Pipino2008}
{Pipino} A., {D'Ercole} A., {Matteucci} F., 2008, \aap, 484, 679

\bibitem[{{Planck Collaboration} {et~al}\mbox{.}(2013){Planck Collaboration},
  {Ade}, {Aghanim}, {Armitage-Caplan}, {Arnaud}, {Ashdown}, {Atrio-Barandela},
  {Aumont}, {Baccigalupi}, {Banday}, \& et~al.}]{PlanckCollaboration2013}
{Planck Collaboration} {et~al.}, 2013, ArXiv e-prints, arXiv:1303.5076

\bibitem[{{Pogge} {et~al}\mbox{.}(2000){Pogge}, {Maoz}, {Ho}, \&
  {Eracleous}}]{Pogge2000}
{Pogge} R.~W., {Maoz} D., {Ho} L.~C., {Eracleous} M., 2000, \apj, 532, 323

\bibitem[{{Riechers} {et~al}\mbox{.}(2013){Riechers}, {Bradford}, {Clements},
  {Dowell}, {P{\'e}rez-Fournon}, {Ivison}, {Bridge}, {Conley}, {Fu}, {Vieira},
  {Wardlow}, {Calanog}, {Cooray}, {Hurley}, {Neri}, {Kamenetzky}, {Aguirre},
  {Altieri}, {Arumugam}, {Benford}, {B{\'e}thermin}, {Bock}, {Burgarella},
  {Cabrera-Lavers}, {Chapman}, {Cox}, {Dunlop}, {Earle}, {Farrah}, {Ferrero},
  {Franceschini}, {Gavazzi}, {Glenn}, {Solares}, {Gurwell}, {Halpern},
  {Hatziminaoglou}, {Hyde}, {Ibar}, {Kov{\'a}cs}, {Krips}, {Lupu}, {Maloney},
  {Martinez-Navajas}, {Matsuhara}, {Murphy}, {Naylor}, {Nguyen}, {Oliver},
  {Omont}, {Page}, {Petitpas}, {Rangwala}, {Roseboom}, {Scott}, {Smith},
  {Staguhn}, {Streblyanska}, {Thomson}, {Valtchanov}, {Viero}, {Wang},
  {Zemcov}, \& {Zmuidzinas}}]{Riechers2013}
{Riechers} D.~A. {et~al.}, 2013, \nat, 496, 329

\bibitem[{{Roediger} {et~al}\mbox{.}(2011){Roediger}, {Courteau}, {MacArthur},
  \& {McDonald}}]{Roediger2011}
{Roediger} J.~C., {Courteau} S., {MacArthur} L.~A., {McDonald} M., 2011,
  \mnras, 416, 1996

\bibitem[{{Salpeter}(1955)}]{Salpeter1955}
{Salpeter} E.~E., 1955, \apj, 121, 161

\bibitem[{{S{\'a}nchez-Bl{\'a}zquez}
  {et~al}\mbox{.}(2007){S{\'a}nchez-Bl{\'a}zquez}, {Forbes}, {Strader},
  {Brodie}, \& {Proctor}}]{Sanchez-Blazquez2007}
{S{\'a}nchez-Bl{\'a}zquez} P., {Forbes} D.~A., {Strader} J., {Brodie} J.,
  {Proctor} R., 2007, \mnras, 377, 759

\bibitem[{{S{\'a}nchez-Bl{\'a}zquez}, {Gorgas} \&
  {Cardiel}(2006){S{\'a}nchez-Bl{\'a}zquez}, {Gorgas}, \&
  {Cardiel}}]{Sanchez-Blazquez2006}
{S{\'a}nchez-Bl{\'a}zquez} P., {Gorgas} J., {Cardiel} N., 2006, \aap, 457, 823

\bibitem[{{Schiavon}(2007)}]{Schiavon2007}
{Schiavon} R.~P., 2007, \apjs, 171, 146

\bibitem[{{Schlegel}, {Finkbeiner} \& {Davis}(1998){Schlegel}, {Finkbeiner}, \&
  {Davis}}]{Schlegel1998}
{Schlegel} D.~J., {Finkbeiner} D.~P., {Davis} M., 1998, \apj, 500, 525

\bibitem[{{S\'{e}rsic}(1968)}]{Sersic1968}
{S\'{e}rsic} J.~L., 1968, {Atlas de galaxias australes}

\bibitem[{{Shen} {et~al}\mbox{.}(2003){Shen}, {Mo}, {White}, {Blanton},
  {Kauffmann}, {Voges}, {Brinkmann}, \& {Csabai}}]{Shen2003}
{Shen} S., {Mo} H.~J., {White} S.~D.~M., {Blanton} M.~R., {Kauffmann} G.,
  {Voges} W., {Brinkmann} J., {Csabai} I., 2003, \mnras, 343, 978

\bibitem[{{Skrutskie} {et~al}\mbox{.}(2006){Skrutskie}, {Cutri}, {Stiening},
  {Weinberg}, {Schneider}, {Carpenter}, {Beichman}, {Capps}, {Chester},
  {Elias}, {Huchra}, {Liebert}, {Lonsdale}, {Monet}, {Price}, {Seitzer},
  {Jarrett}, {Kirkpatrick}, {Gizis}, {Howard}, {Evans}, {Fowler}, {Fullmer},
  {Hurt}, {Light}, {Kopan}, {Marsh}, {McCallon}, {Tam}, {Van Dyk}, \&
  {Wheelock}}]{Skrutskie2006}
{Skrutskie} M.~F. {et~al.}, 2006, \aj, 131, 1163

\bibitem[{{Sohn} {et~al}\mbox{.}(2006){Sohn}, {O'Connell}, {Kundu}, {Landsman},
  {Burstein}, {Bohlin}, {Frogel}, \& {Rose}}]{Sohn2006}
{Sohn} S.~T., {O'Connell} R.~W., {Kundu} A., {Landsman} W.~B., {Burstein} D.,
  {Bohlin} R.~C., {Frogel} J.~A., {Rose} J.~A., 2006, \aj, 131, 866

\bibitem[{{Spolaor} {et~al}\mbox{.}(2008){Spolaor}, {Forbes}, {Proctor}, {Hau},
  \& {Brough}}]{Spolaor2008}
{Spolaor} M., {Forbes} D.~A., {Proctor} R.~N., {Hau} G.~K.~T., {Brough} S.,
  2008, \mnras, 385, 675

\bibitem[{{Spolaor} {et~al}\mbox{.}(2010){Spolaor}, {Kobayashi}, {Forbes},
  {Couch}, \& {Hau}}]{Spolaor2010}
{Spolaor} M., {Kobayashi} C., {Forbes} D.~A., {Couch} W.~J., {Hau} G.~K.~T.,
  2010, \mnras, 408, 272

\bibitem[{{Spolaor} {et~al}\mbox{.}(2009){Spolaor}, {Proctor}, {Forbes}, \&
  {Couch}}]{Spolaor2009}
{Spolaor} M., {Proctor} R.~N., {Forbes} D.~A., {Couch} W.~J., 2009, \apjl, 691,
  L138

\bibitem[{{Tal} \& {van Dokkum}(2011)}]{Tal2011}
{Tal} T., {van Dokkum} P.~G., 2011, \apj, 731, 89

\bibitem[{{Thomas} {et~al}\mbox{.}(2005){Thomas}, {Maraston}, {Bender}, \&
  {Mendes de Oliveira}}]{Thomas2005}
{Thomas} D., {Maraston} C., {Bender} R., {Mendes de Oliveira} C., 2005, \apj,
  621, 673

\bibitem[{{Trujillo} {et~al}\mbox{.}(2013){Trujillo}, {Ferre-Mateu},
  {Balcells}, {Vazdekis}, \& {Sanchez-Blazquez}}]{Trujillo2013}
{Trujillo} I., {Ferre-Mateu} A., {Balcells} M., {Vazdekis} A.,
  {Sanchez-Blazquez} P., 2013, ArXiv e-prints

\bibitem[{{Trujillo}, {Ferreras} \& {de La Rosa}(2011){Trujillo}, {Ferreras},
  \& {de La Rosa}}]{Trujillo2011}
{Trujillo} I., {Ferreras} I., {de La Rosa} I.~G., 2011, \mnras, 415, 3903

\bibitem[{{Trujillo} {et~al}\mbox{.}(2006){Trujillo}, {F{\"o}rster Schreiber},
  {Rudnick}, {Barden}, {Franx}, {Rix}, {Caldwell}, {McIntosh}, {Toft},
  {H{\"a}ussler}, {Zirm}, {van Dokkum}, {Labb{\'e}}, {Moorwood},
  {R{\"o}ttgering}, {van der Wel}, {van der Werf}, \& {van
  Starkenburg}}]{Trujillo2006}
{Trujillo} I. {et~al.}, 2006, \apj, 650, 18

\bibitem[{{Trujillo}, {Graham} \& {Caon}(2001){Trujillo}, {Graham}, \&
  {Caon}}]{Trujillo2001}
{Trujillo} I., {Graham} A.~W., {Caon} N., 2001, \mnras, 326, 869

\bibitem[{{van Dokkum} \& {Conroy}(2010)}]{vanDokkum2010}
{van Dokkum} P.~G., {Conroy} C., 2010, \nat, 468, 940

\bibitem[{{van Dokkum} {et~al}\mbox{.}(2010){van Dokkum}, {Whitaker},
  {Brammer}, {Franx}, {Kriek}, {Labb{\'e}}, {Marchesini}, {Quadri}, {Bezanson},
  {Illingworth}, {Muzzin}, {Rudnick}, {Tal}, \& {Wake}}]{vanDokkum2010a}
{van Dokkum} P.~G. {et~al.}, 2010, \apj, 709, 1018

\bibitem[{{White}(1980)}]{White1980}
{White} S.~D.~M., 1980, \mnras, 191, 1P

\bibitem[{{York} {et~al}\mbox{.}(2000){York}, {Adelman}, {Anderson},
  {Anderson}, {Annis}, {Bahcall}, {Bakken}, {Barkhouser}, {Bastian}, {Berman},
  {Boroski}, {Bracker}, {Briegel}, {Briggs}, {Brinkmann}, {Brunner}, {Burles},
  {Carey}, {Carr}, {Castander}, {Chen}, {Colestock}, {Connolly}, {Crocker},
  {Csabai}, {Czarapata}, {Davis}, {Doi}, {Dombeck}, {Eisenstein}, {Ellman},
  {Elms}, {Evans}, {Fan}, {Federwitz}, {Fiscelli}, {Friedman}, {Frieman},
  {Fukugita}, {Gillespie}, {Gunn}, {Gurbani}, {de Haas}, {Haldeman}, {Harris},
  {Hayes}, {Heckman}, {Hennessy}, {Hindsley}, {Holm}, {Holmgren}, {Huang},
  {Hull}, {Husby}, {Ichikawa}, {Ichikawa}, {Ivezi{\'c}}, {Kent}, {Kim},
  {Kinney}, {Klaene}, {Kleinman}, {Kleinman}, {Knapp}, {Korienek}, {Kron},
  {Kunszt}, {Lamb}, {Lee}, {Leger}, {Limmongkol}, {Lindenmeyer}, {Long},
  {Loomis}, {Loveday}, {Lucinio}, {Lupton}, {MacKinnon}, {Mannery}, {Mantsch},
  {Margon}, {McGehee}, {McKay}, {Meiksin}, {Merelli}, {Monet}, {Munn},
  {Narayanan}, {Nash}, {Neilsen}, {Neswold}, {Newberg}, {Nichol}, {Nicinski},
  {Nonino}, {Okada}, {Okamura}, {Ostriker}, {Owen}, {Pauls}, {Peoples},
  {Peterson}, {Petravick}, {Pier}, {Pope}, {Pordes}, {Prosapio},
  {Rechenmacher}, {Quinn}, {Richards}, {Richmond}, {Rivetta}, {Rockosi},
  {Ruthmansdorfer}, {Sandford}, {Schlegel}, {Schneider}, {Sekiguchi}, {Sergey},
  {Shimasaku}, {Siegmund}, {Smee}, {Smith}, {Snedden}, {Stone}, {Stoughton},
  {Strauss}, {Stubbs}, {SubbaRao}, {Szalay}, {Szapudi}, {Szokoly}, {Thakar},
  {Tremonti}, {Tucker}, {Uomoto}, {Vanden Berk}, {Vogeley}, {Waddell}, {Wang},
  {Watanabe}, {Weinberg}, {Yanny}, {Yasuda}, \& {SDSS
  Collaboration}}]{York2000}
{York} D.~G. {et~al.}, 2000, \aj, 120, 1579

\bibitem[{{Zibetti} {et~al}\mbox{.}(2013){Zibetti}, {Gallazzi}, {Charlot},
  {Pierini}, \& {Pasquali}}]{Zibetti2013}
{Zibetti} S., {Gallazzi} A., {Charlot} S., {Pierini} D., {Pasquali} A., 2013,
  \mnras, 428, 1479

\end{thebibliography}
\bibliographystyle{mn2e}
\bsp
\appendix

\section[]{The Radial Profiles of the Centre of M$87$}

Here we present the photometric data derived for the centre of M$87$ using HR imaging. The photometric errors are derived from the residuals to S\'{e}rsic fits as explained in Section \ref{errors}. The quality of STIS F25QTZ image (NUV) just allows us to derive reliable photometry up to R$=5.5 \arcsec$. 

\tiny
\setlongtables
\begin{landscape}
\onecolumn
 \begin{longtable}{ccccccccccccccccc} 
  \caption{Data for the inner $19\arcsec$ of M$87$} \\
   Radius  & FUV & NUV &  F336W & F410M & F467M & F475W & F555W & F547M & F606W & F658N & F702W & F814W & F850LP & $J$ & F$160$W & $K_s$\\ 
 ($\arcsec$) &  \multicolumn{16}{c}{\cellcolor[gray]{0.8}[mag/arcsec$^2$]}\\\hline
 
 $ 1.5$ & $ 21.93 $ & $ 21.66 $ & $ 19.76 $ & $ 18.50 $ & $ 17.67 $ & $ 17.63 $ & $ 17.15 $ & $ 17.03 $ & $ 16.85 $ & $ 16.45 $ & $ 16.48 $ & $ 16.18 $ & $ 15.97 $ & $ 15.61 $ & $ 15.33 $ & $ 15.54    $    \\
 $ 2.5$ & $ 22.15 $ & $ 21.96 $ & $ 19.94 $ & $ 18.64 $ & $ 17.81 $ & $ 17.77 $ & $ 17.30 $ & $ 17.17 $ & $ 17.00 $ & $ 16.67 $ & $ 16.64 $ & $ 16.33 $ & $ 16.11 $ & $ 15.74 $ & $ 15.48 $ & $ 15.69    $    \\
 $ 3.5$ & $ 22.32 $ & $ 22.16 $ & $ 20.07 $ & $ 18.76 $ & $ 17.93 $ & $ 17.89 $ & $ 17.41 $ & $ 17.29 $ & $ 17.13 $ & $ 16.86 $ & $ 16.76 $ & $ 16.45 $ & $ 16.24 $ & $ 15.86 $ & $ 15.61 $ & $ 15.82    $    \\
 $ 4.5$ & $ 22.47 $ & $ 22.34 $ & $ 20.19 $ & $ 18.87 $ & $ 18.05 $ & $ 18.00 $ & $ 17.53 $ & $ 17.40 $ & $ 17.25 $ & $ 17.00 $ & $ 16.88 $ & $ 16.57 $ & $ 16.35 $ & $ 15.97 $ & $ 15.72 $ & $ 15.94    $    \\
 $ 5.5$ & $ 22.60 $ & $ 22.53 $ & $ 20.31 $ & $ 18.98 $ & $ 18.16 $ & $ 18.12 $ & $ 17.65 $ & $ 17.52 $ & $ 17.37 $ & $ 17.12 $ & $ 17.00 $ & $ 16.69 $ & $ 16.48 $ & $ 16.08 $ & $ 15.84 $ & $ 16.06    $    \\
 $ 6.5$ & $ 22.75 $ & $   ... $ & $ 20.43 $ & $ 19.10 $ & $ 18.29 $ & $ 18.24 $ & $ 17.77 $ & $ 17.64 $ & $ 17.49 $ & $ 17.25 $ & $ 17.12 $ & $ 16.81 $ & $ 16.60 $ & $ 16.20 $ & $ 15.96 $ & $ 16.18    $    \\
 $ 7.5$ & $ 22.89 $ & $   ... $ & $ 20.55 $ & $ 19.23 $ & $ 18.41 $ & $ 18.37 $ & $ 17.89 $ & $ 17.77 $ & $ 17.62 $ & $ 17.37 $ & $ 17.25 $ & $ 16.94 $ & $ 16.72 $ & $ 16.31 $ & $ 16.08 $ & $ 16.30    $    \\
 $ 8.5$ & $ 22.99 $ & $   ... $ & $ 20.68 $ & $ 19.35 $ & $ 18.54 $ & $ 18.49 $ & $ 18.01 $ & $ 17.89 $ & $ 17.74 $ & $ 17.50 $ & $ 17.37 $ & $ 17.07 $ & $ 16.85 $ & $ 16.42 $ & $ 16.20 $ & $ 16.43    $    \\
 $ 9.5$ & $ 23.14 $ & $   ... $ & $ 20.80 $ & $ 19.47 $ & $ 18.66 $ & $ 18.61 $ & $ 18.13 $ & $ 18.02 $ & $ 17.86 $ & $ 17.62 $ & $ 17.49 $ & $ 17.19 $ & $ 16.97 $ & $ 16.52 $ & $ 16.32 $ & $ 16.55    $    \\
 $10.5$ & $ 23.25 $ & $   ... $ & $ 20.93 $ & $ 19.59 $ & $ 18.78 $ & $ 18.73 $ & $ 18.25 $ & $ 18.14 $ & $ 17.98 $ & $ 17.75 $ & $ 17.62 $ & $ 17.31 $ & $ 17.09 $ & $   ... $ & $   ... $ & $   ...    $    \\
 $11.5$ & $ 23.36 $ & $   ... $ & $ 21.05 $ & $ 19.70 $ & $ 18.90 $ & $ 18.84 $ & $ 18.37 $ & $ 18.26 $ & $ 18.10 $ & $ 17.86 $ & $ 17.74 $ & $ 17.43 $ & $ 17.21 $ & $   ... $ & $   ... $ & $   ...    $    \\
 $12.5$ & $ 23.44 $ & $   ... $ & $ 21.17 $ & $ 19.81 $ & $ 19.02 $ & $ 18.95 $ & $ 18.48 $ & $ 18.38 $ & $ 18.21 $ & $ 17.97 $ & $ 17.85 $ & $ 17.54 $ & $ 17.32 $ & $   ... $ & $   ... $ & $   ...    $    \\
 $13.5$ & $ 23.54 $ & $   ... $ & $ 21.29 $ & $ 19.92 $ & $ 19.13 $ & $ 19.06 $ & $ 18.59 $ & $ 18.49 $ & $ 18.32 $ & $ 18.08 $ & $ 17.96 $ & $ 17.65 $ & $ 17.43 $ & $   ... $ & $   ... $ & $   ...    $    \\
 $14.5$ & $ 23.62 $ & $   ... $ & $ 21.40 $ & $ 20.02 $ & $ 19.24 $ & $ 19.16 $ & $ 18.69 $ & $ 18.59 $ & $ 18.42 $ & $ 18.19 $ & $ 18.07 $ & $ 17.75 $ & $ 17.53 $ & $   ... $ & $   ... $ & $   ...    $    \\
 $15.5$ & $ 23.71 $ & $   ... $ & $ 21.50 $ & $ 20.11 $ & $ 19.35 $ & $ 19.25 $ & $ 18.79 $ & $ 18.69 $ & $ 18.52 $ & $ 18.30 $ & $ 18.17 $ & $ 17.85 $ & $ 17.63 $ & $   ... $ & $   ... $ & $   ...    $    \\
 $16.5$ & $ 23.78 $ & $   ... $ & $ 21.57 $ & $ 20.21 $ & $ 19.43 $ & $ 19.35 $ & $ 18.89 $ & $ 18.78 $ & $ 18.61 $ & $ 18.40 $ & $ 18.27 $ & $ 17.94 $ & $ 17.73 $ & $   ... $ & $   ... $ & $   ...    $    \\
 $17.5$ & $ 23.85 $ & $   ... $ & $ 21.64 $ & $ 20.30 $ & $ 19.51 $ & $ 19.43 $ & $ 18.99 $ & $ 18.87 $ & $ 18.70 $ & $ 18.49 $ & $ 18.37 $ & $ 18.04 $ & $ 17.82 $ & $   ... $ & $   ... $ & $   ...    $    \\
\label{tablarads} \\
\hline
 \end{longtable}
\end{landscape}
 \twocolumn
\normalsize
 
\section[]{S\'ersic fits residuals and HR/wide-field offsets}\label{appendixa}

\begin{figure}
  \begin{center}
   \includegraphics[height=0.45\textheight]{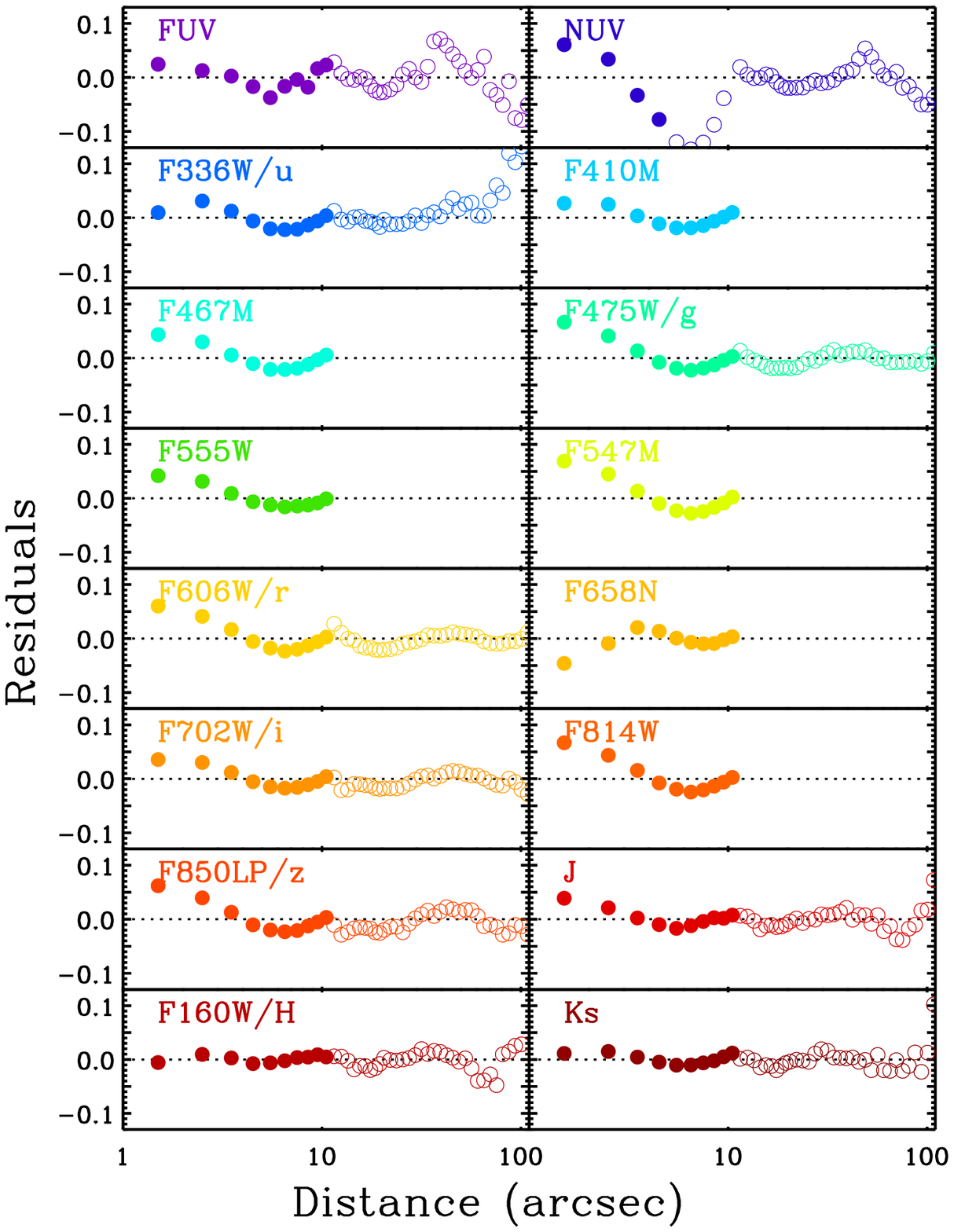} 
  \caption{Residuals of the observed profiles after substracting their corresponding S\'ersic fits. The filled circles indicate HR profiles while the open circles depict the wide-fiel data.}\label{app:sersic}
 \end{center}
\end{figure}

To illustrate how the errors have been estimated, we provide the residuals of the profiles to the corresponding S\'ersic fits for HR and wide-field data in Figure \ref{app:sersic}. In Table \ref{app:tablashift}, we also reported the shifts and the root mean square (r.m.s) between HR and wide-field profiles in the overlapping region.

\begin{table}
 \begin{center}
  \begin{tabular}{ccc} \\
   Filter        & Shift & r.m.s \\ \hline 
  \multicolumn{3}{c}{\cellcolor[gray]{0.8}[mag/arcsec$^2$]}\\\hline
   FUV STIS & -0.21 &  0.096 \\
   NUV STIS & 0.50  &        \\
   F336W & -0.25& 0.050\\
   F410M & -0.02& 0.020\\
   F467M & -0.02& 0.010\\
   F475W & -0.04&  0.006  \\
   F555W & -0.10& 0.010\\
   F547M & -0.40& 0.030\\
   F606W & -0.11& 0.007 \\
   F702W & -0.15& 0.014\\
   F814W & -0.10& 0.009\\
   F850LP & -0.10&  0.015  \\
   J NaCo & 0.08 &  0.014  \\
   F160W  & -0.01&  0.005  \\
   K NaCo & 0.0  &  0.010  \\ \hline
  \end{tabular}\caption{Offset and r.m.s between HR and wide-field profiles.}\label{app:tablashift}
 \end{center}
\end{table}

\section[]{Age and metallicity radial profiles for different constraints}\label{appendixb}

\begin{figure}
 \begin{center}
  \includegraphics[height=0.3\textheight]{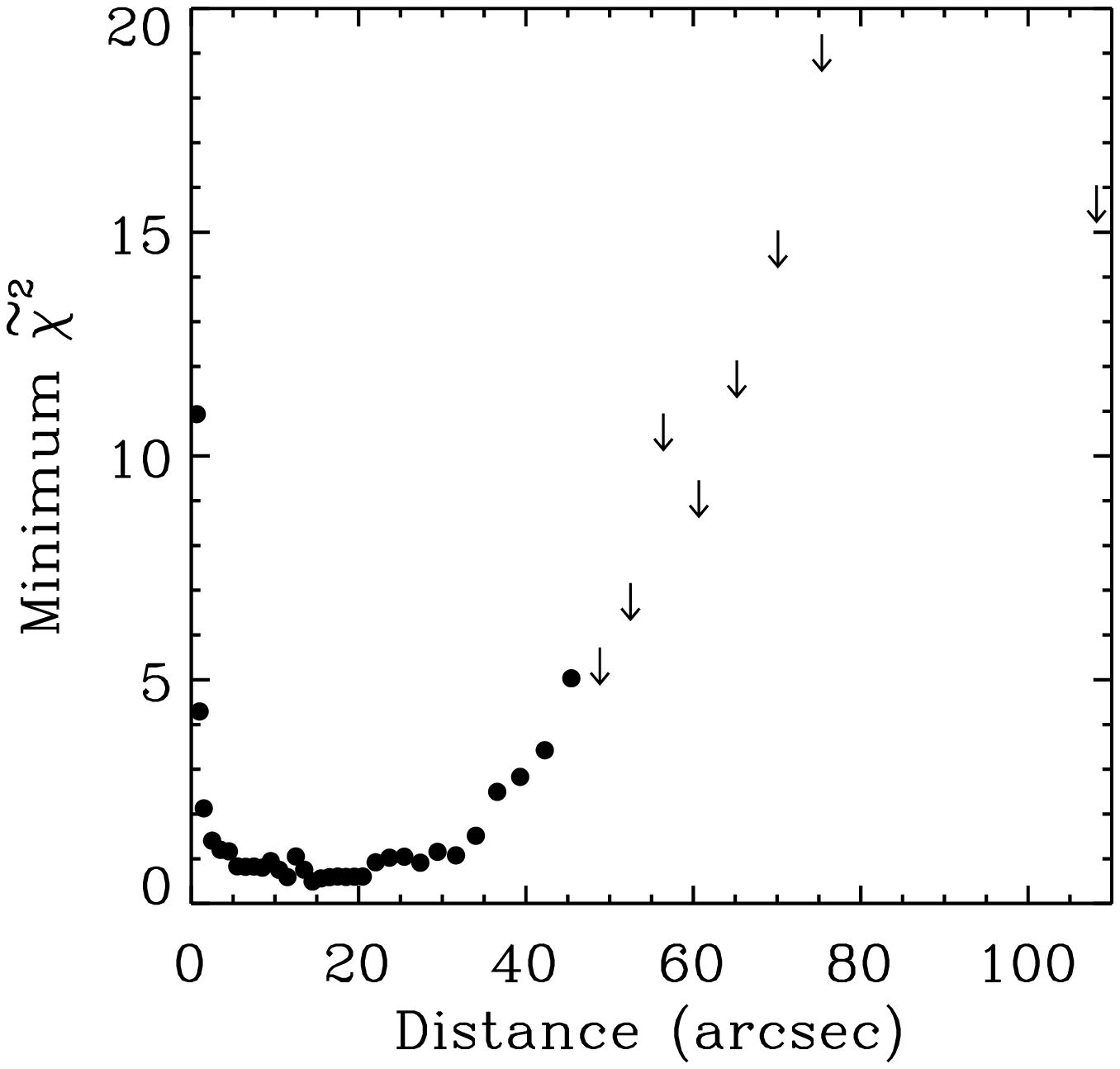} 
   \caption{Minimum $\tilde{\chi}^2$ as a function of radius. It is clearly seen the effect of the incorrect sky substraction for $u$ and 2MASS photometry at R$>40\arcsec$.}\label{app:minchi}
 \end{center}
\end{figure}

Here we present the results of the fits to the radial SEDs of M87, following Section \ref{SSPfitting}. Figure \ref{app:minchi} shows the minimum $\tilde{\chi}^2$ as a function of radius corresponding to Figure \ref{ageandmet}. When the model is a good representation of the data and the errors are properly estimated, the expected values of $\tilde{\chi}^2$ should be around $1$. For R$\gtrsim40\arcsec$, the increase in the value of the $\tilde{\chi}^2$ is due to the effect of the sky oversubstraction in the 2MASS and $u$ profiles. For this reason, the values are marked as upper limits.

\begin{figure*}
 \begin{center}
   \includegraphics[height=0.3\textheight]{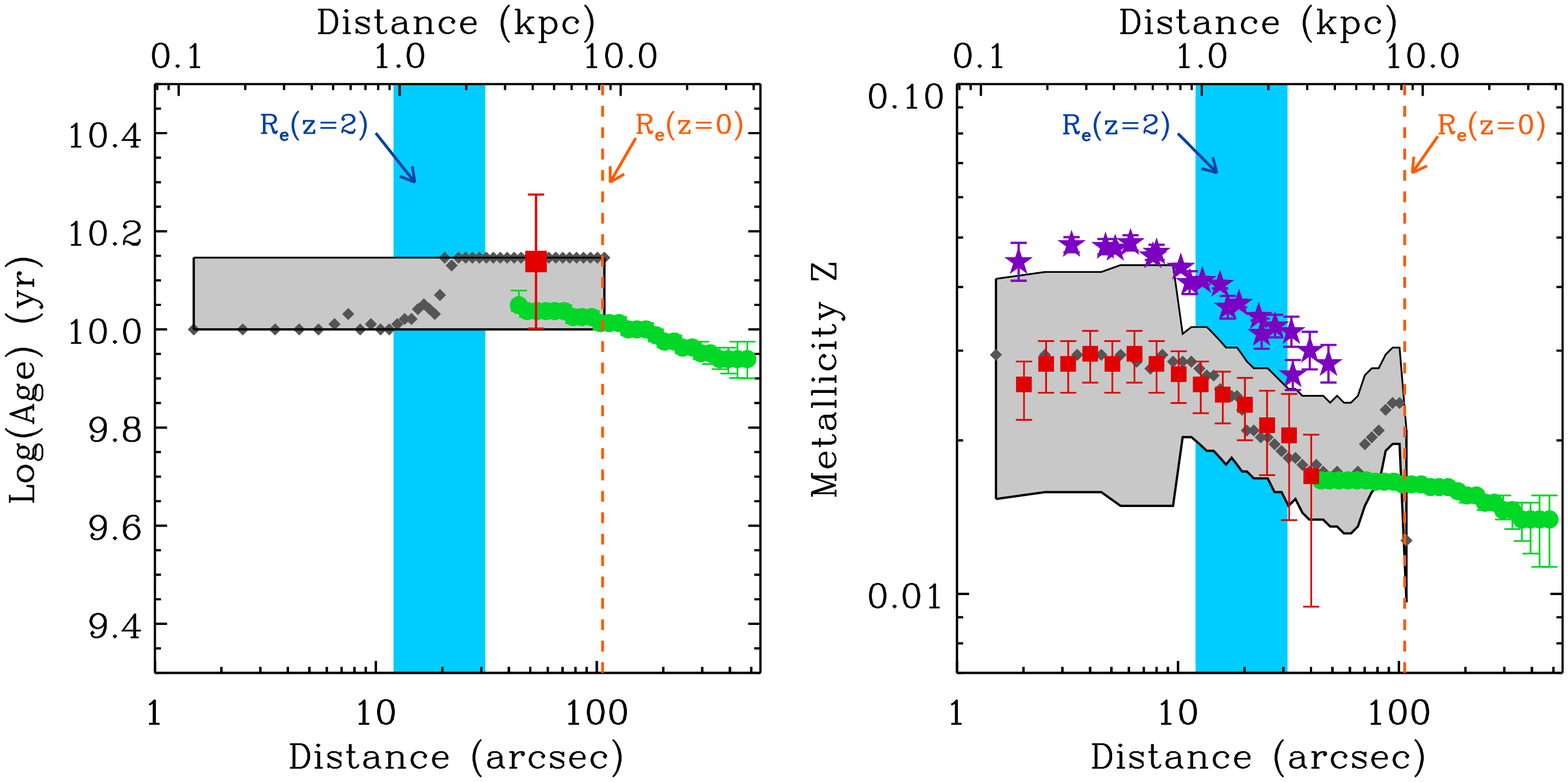} 
  \caption{The same as in Figure \ref{ageandmet} but using the BC03 SSP models with ages constrained from $10$ to $14$ Gyr. The gradient is: $-0.27\pm0.08$. }\label{app:ageandmet}
 \end{center}
\end{figure*}

\begin{figure*}
 \begin{center}
   \includegraphics[height=0.3\textheight]{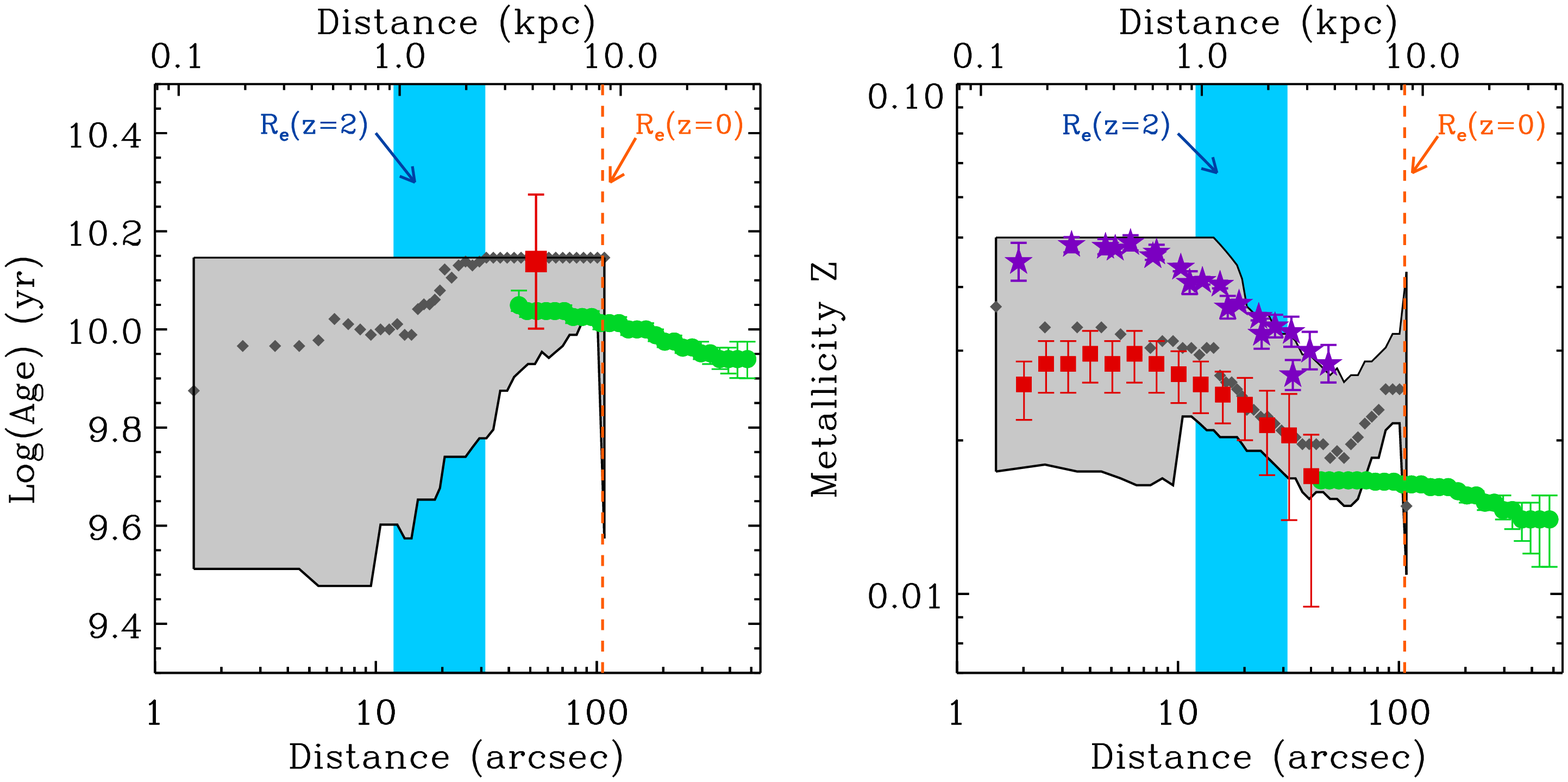} 
  \caption{The same as in Figure \ref{ageandmet} but using the BC03 SSP model with a \citet{Chabrier2003} IMF. The gradient is: $-0.25\pm0.09$. }\label{app:ageandmet_chab}
 \end{center}
\end{figure*}

\begin{figure*}
 \begin{center}
  \includegraphics[height=0.3\textheight]{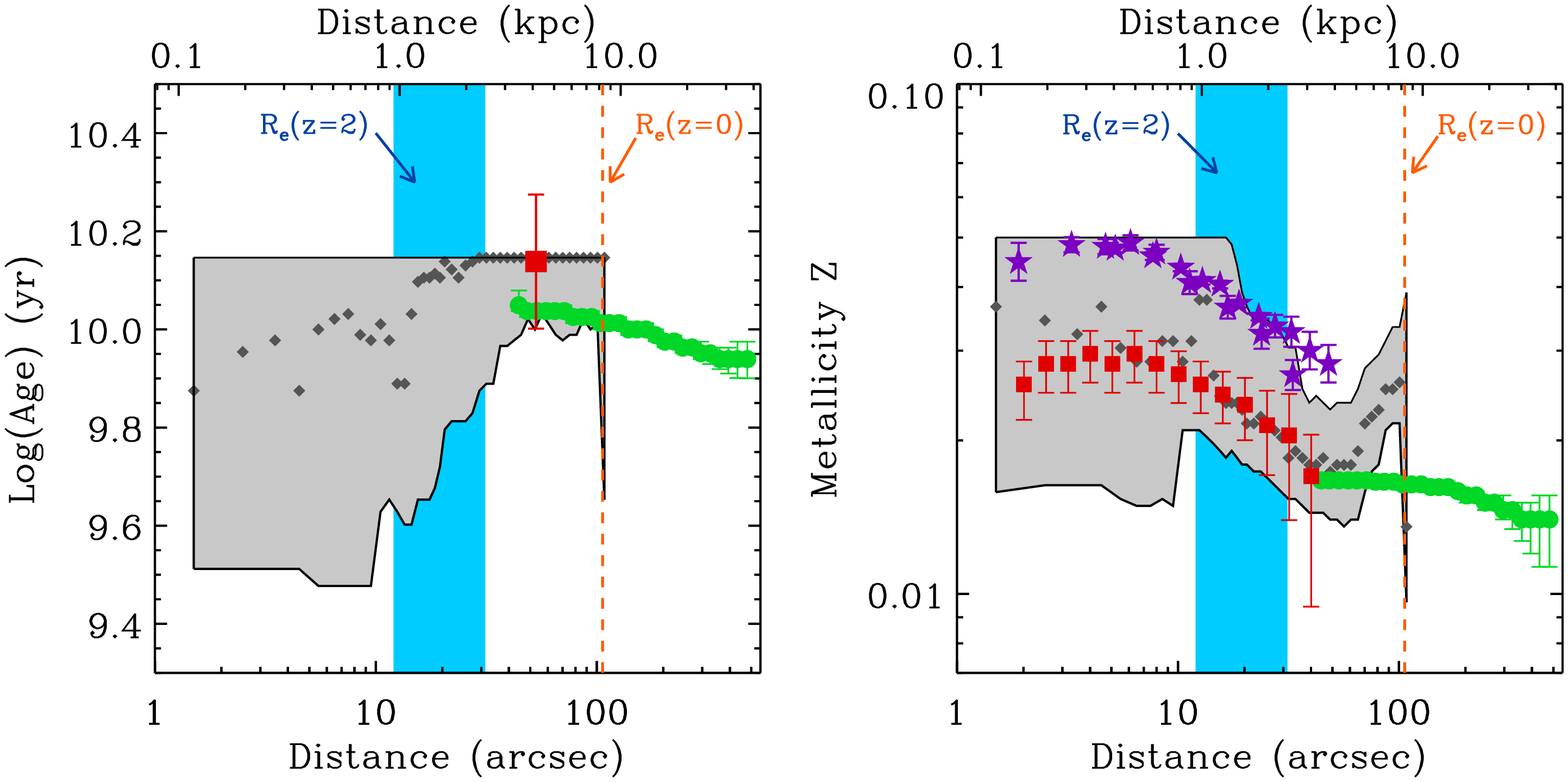} 
  \caption{The same as in Figure \ref{ageandmet} but using the Charlot \& Bruzual (2007, in prep.) models. The gradient in this case is: $-0.29\pm0.09$. }\label{app:ageandmet_07}
 \end{center}
\end{figure*}

We have also tested different SSPs contraints and models to check the reliability of our results. First, we imposed an age constraint: from $10$ to $14$ Gyr. It seems reasonable to assume an old age for the bulk of stars of the galaxy given the results of previous studies \citep[e.g.][]{Liu2005,Kuntschner2010}. The results are presented in Figure \ref{app:ageandmet}. The uncertainty on the metallicity profile has decreased in the inner $20\arcsec$, not reaching the highest metallicity available for the models. Consequently, this result supports the idea that the \textit{plateau} at R$<1$ kpc is a real characteristic of the metallicity profile of M87 and not an artifact due to the limitations of the models. We also present the results of using: BC03 models with a \citet{Chabrier2003} in Figure \ref{app:ageandmet_chab} and Charlot \& Bruzual (2007, priv. comm.) in Figure \ref{app:ageandmet_07}. The gradients are: $-0.27\pm0.08$, $-0.25\pm0.09$ and $-0.29\pm0.09$, respectively.

\label{lastpage}

\end{document}